\documentclass[journal]{IEEEtran} %
\usepackage{cite}
\usepackage{framed}
\usepackage{color}
\definecolor{shadecolor}{gray}{0.90}

\ifCLASSINFOpdf
\else
  \usepackage[dvips]{graphicx}
\fi
\usepackage{amssymb,bm,mathrsfs,bbm,amscd,amsmath}

\newtheorem{Remark}{Remark}
\usepackage{paralist}

\usepackage{subfig}
\usepackage{tabularx}
\usepackage{multirow}
\usepackage[ruled]{algorithm2e}  

\SetKwInput{KwData}{\textbf{Initialization}}
\usepackage{verbatim}
\usepackage{hyperref}
\hypersetup{hidelinks}

\usepackage{color}

\begin{document}
%
\title{Reliable OFDM Receiver with \\ Ultra-Low Resolution ADC}
%
%
%

\author{Hanqing~Wang,~\IEEEmembership{Student~Member,~IEEE,}
	    Wan-Ting Shih,~
        Chao-Kai~Wen,~\IEEEmembership{Member,~IEEE,}\\
        and~Shi~Jin,~\IEEEmembership{Senior Member,~IEEE}
\thanks{H. Wang and S. Jin are with the National Mobile Communications Research Laboratory, Southeast University, Nanjing 210096, P. R. China. P (e-mail: $\rm hqwanglyt@seu.edu.cn;~jinshi@seu.edu.cn$).}
\thanks{W.-T. Shih is with the Institute of Electrical and Control Engineering,
National Chiao Tung University, Hsinchu 30010, Taiwan (e-mail: $\rm sydney2317076@gmail.com$).}
\thanks{C.-K. Wen is with the Institute of Communications Engineering, National Sun Yat-sen University, Kaohsiung, Taiwan (e-mail: $\rm chaokai.wen@mail.nsysu.edu.tw$).}
}

\maketitle

\begin{abstract}
	The use of low-resolution analog-to-digital converters (ADCs) can significantly reduce power consumption and hardware cost.
	However, their resulting severe nonlinear distortion makes reliable data transmission challenging.
	For orthogonal frequency division multiplexing (OFDM) transmission, the orthogonality among subcarriers is destroyed.
	This invalidates conventional OFDM receivers relying heavily on this orthogonality.
	In this study, we move on to quantized OFDM (Q-OFDM) prototyping implementation based on our previous achievement in optimal Q-OFDM detection.
	First, we propose a novel Q-OFDM channel estimator by extending the generalized Turbo (GTurbo) framework formerly applied for optimal detection.
	Specifically, we integrate a type of robust linear OFDM channel estimator into the original GTurbo framework and derive its corresponding extrinsic information to guarantee its convergence.
	We also propose feasible schemes for automatic gain control, noise power estimation, and synchronization.
	Combined with the proposed inference algorithms, we develop an efficient Q-OFDM receiver architecture.
	Furthermore, we construct a proof-of-concept prototyping system and conduct over-the-air (OTA) experiments to examine its feasibility and reliability.
	This is the first work that focuses on both algorithm design and system implementation in the field of low-resolution quantization communication.
	The results of the numerical simulation and OTA experiment demonstrate that reliable data transmission can be achieved. 
\end{abstract}


%
\IEEEpeerreviewmaketitle

\section{Introduction}
An essential pursuit of current research into future wireless communication systems is to achieve highly reliable data transmission using devices with ultra-low cost and power consumption. Low-resolution analog-to-digital converter (ADC) is the most representative one, whose application in communication receivers has been extensively explored by the academia and industry.
This emphasis on low-resolution ADCs is motivated by their obvious advantages in reducing hardware cost and power consumption.
Specifically, according to the power consumption expression of ADCs given by \cite{Walden-1999JSAC},  
\[{{P_{{\rm{ADC}}}} \propto {2^b} \times {F_s}},\]
where $b$ denotes its quantization precision and $F_s$ denotes its sampling frequency,
the reduction in the quantization precision $b$ leads to an exponential decrease in the power consumption of ADCs.
The hardware cost and implementation complexity of low-resolution ADCs are also relatively low \cite{ADC_Survey}.
Furthermore, storing and transferring data digitally represented by fewer bits are easier,
which relaxes the quality requirement for radio frequency (RF) and interface circuits, and also reduces the power consumption of these components.
 
The feasibility of using low-resolution ADCs has been initially demonstrated by studies on the capacity or achievable rate degradation caused by such devices \cite{Nossek-2008ISIT,Madhow-2009TCOM,Krone-2010ITW,RHeath-2015TSP,Liang-2016TCOM,LFan-2015CL,Eldar-2017Arxiv}.
As indicated by the capacity derivation for additive white Gaussian noise (AWGN) channels in \cite{Madhow-2009TCOM}, only 10\%-20\% of capacity is lost under the quantization precision of as few as 2$-$3 bits.
The research in \cite{Liang-2016TCOM} revealed that in multiple-input and multiple-output (MIMO) systems with mixed-precision ADCs, equipping only a small proportion of high-resolution ADCs can closely approximate the achievable rate of conventional architecture with all full-resolution ADCs.
These results inspired subsequent studies into signal recovery techniques for dealing with strong nonlinear distortion caused by coarse quantization.
Various methods have been proposed. They include linear receiver \cite{Larsson-2014ArXiv,Nossek-2007WSA, RHeath-2016ICASSP}, projected gradient	method based algorithms \cite{RHeath-2016TCOM,Alevizos-2018CL}, fast adaptive shrinkage/thresholding algorithm (FASTA) based method \cite{Studer-2016TCOM}, coding theory-based approach \cite{NLee-2017Arxiv1}, supervised learning based method \cite{NLee-2017Arxiv2}, and message-passing approaches, including generalized approximate message passing (GAMP) \cite{Wang-2014ICC, Mezghani-2010ISIT,TJiang-2016TWC,Nossek-2016ArXiv} and its extension bilinear GAMP (BiGAMP) \cite{CKWen-2016TSP} and parametric BiGAMP \cite{Heath-2018Arxiv}, vector AMP \cite{Schniter-2016Arxiv}, and generalized Turbo (GTurbo) \cite{Wang-2017JSAC}.

In this study, we focus on orthogonal frequency division multiplexing (OFDM) systems with ultra-low resolution ADCs (i.e., 1$-$2 bits) at the receiver.
We refer to these systems as quantized OFDM (Q-OFDM) systems.
OFDM is a well-developed multicarrier technology for decomposing convolutive multipath channels into a bank of parallel flat-fading subchannels to facilitate wideband transmission with high spectral efficiency \cite{Proakis2007}.
Hence, OFDM is applied as the physical layer waveform for variety of modern high-rate wireless systems, ranging from Long-Term Evolution and its evolutions to the standards such as IEEE 802.11ad \cite{r1}.
It is also one of the strong waveform contenders for the 5th generation (5G) air interface \cite{Schaich-2014ISCCSP,Zaidi-2016MCOM}.
This research considers the most fundamental single-stream transmission case. 
Using ultra-low-resolution quantizers results in only minimal performance loss when many received streams, which average the quantization noise, are available for combination  \cite{Larsson-2014ArXiv}. 
However, in case of few received streams, in which diversity gain from a large number of received streams is unavailable, reliable reception is a challenging task in the presence of coarse quantization.
Single stream is considered as an extreme case.
In addition, this single stream input-output relation does not solely imply the single-antenna case. 
It can also be treated as an equivalent representation for multi-antenna systems with one transmitted and received data stream using analog beamforming when tackling data detection and channel estimation, as applied in \cite{Yin-13GLCOM,Heath-2018Arxiv}.

The conventional OFDM receiver proposed for the high-resolution quantization case performs by converting the time-domain received signal into the frequency-domain. 
The frequency-domain observations can be regarded as a bank of orthogonal flat-fading subchannels or subcarriers. 
Data detection and channel estimation are then implemented in parallel for each flat-fading subchannel. 
However, in the presence of coarse quantization, observations among subcarriers are not orthogonal anymore.
Moreover, high correlations are exhibited in the frequency domain, and severe inter-carrier interference occurs. 
This conventional strategy, which is derived from orthogonality, can suffer from certain performance loss for Q-OFDM cases. 
Such occurrences have motivated us to propose advanced receiver strategies.
Our pioneering work \cite{Wang-2017JSAC} proposed a powerful Q-OFDM detector based on the GTurbo signal recovery principle.
This detector proved theoretically optimal in the sense of minimum mean-square error (MMSE) with ideal channel state information (CSI).
In addition, an efficient pipelined and folding hardware architecture of the GTurbo algorithm was proposed in \cite{CZhang-2016SiPS}, which suggests that hardware friendly implementation of GTurbo-based algorithms is achievable.

In this work, we move on to the next step to realize the Q-OFDM prototyping system for reliable data transmission.
To this end, several issues are waiting for being tackled.
One important issue is channel estimation as frequency-domain channel responses are necessary for the Q-OFDM detector.
Considering the algorithmic power of GTurbo in Q-OFDM detection, we develop a novel Q-OFDM channel estimator that is based on the GTurbo architecture. 
The main difference or challenge is that accurately characterizing the prior distribution of frequency-domain channel responses, which are time-variant and correlated, is difficult. 
Then the MMSE estimate of frequency-domain channel responses required in the original GTurbo framework is infeasible. 
To tackle this issue, we embed a type of robust OFDM channel estimator used for OFDM systems with high-resolution quantization into the GTurbo framework. 
In this way, our approach can be viewed as performing linear channel estimation followed by a certain linear approximation that is updated via iterations. 
As shown by the numerical results, the proposed scheme outperforms its counterparts, such as the conventional method, FASTA \cite{Studer-2016TCOM}, expectation-maximization Gaussian-mixture GAMP (EM-GM-GAMP) \cite{Vila-2013TSP}, and Bussgang's theorem-based method \cite{Bussgang52}. 
In addition, the developments of detection and channel estimation based on the same algorithm architecture allow for sharing an identical circuit module when designing specific circuits for the proposed algorithms.

For prototyping implementation, we also need to tackle several critical issues, including adjusting the input signal amplitude within a certain range (i.e., automatic gain control (AGC)), estimating the noise power, and identifying the starting point of each radio frame. AGC and noise estimation require the amplitude information of the received signal.
As a low-resolution ADC eliminates its amplitude information, we introduce another high-resolution ADC with a considerably low sampling rate.
With a power accumulator, signal and noise power can be measured.
Moreover, we verify that correlation-based synchronization works well for coarsely quantized received signals.
Combining the above issues with the proposed GTurbo-based algorithms, we put forward an efficient Q-OFDM receiver architecture and construct a prototyping system on the basis of such architecture.
To the best of our knowledge, our current work is the first one which covers both theory (algorithm design) and practice (prototyping implementation and over-the-air (OTA) experiments) in the field of communication with low-resolution ADCs.

OTA experiments are conducted for justifying the reliability of the proposed architecture. 
Results reveal that reliable OFDM transmissions of 4-QAM symbols with 1-bit quantization per real dimension and 16-QAM symbols with 2-bit quantization per real dimension can be achieved. 
In addition, the well-known guideline for AWGN channels, which claims that the ADC precision per real dimension equal to the logarithm of half  the modulation order is sufficient for reliable signal recovery \cite{Madhow-2009TCOM}, is also valid for OFDM transmissions. 
However, this guideline is not intuitively true for OFDM transmission, given that the inverse discrete Fourier transform (IDFT) operation produces continuous time-domain transmitted symbols instead of discrete constellation points.
Thus, higher ADC precision is required. 
This interesting finding verifies the powerful capacity of the proposed algorithms and receiver framework.

This work, along with \cite{Wang-2017JSAC}, forms a complete picture of the reliable Q-OFDM receiver for the one-stream transmission case. 
These studies help us clarify many essential technical issues, thereby establishing the basis of the future extension to multi-antenna and multi-user cases. 
Specifically, novel algorithm architectures potential for the single-user MIMO case can be developed by extending current GTurbo-based framework, such as \cite{He-2018ISIT}. 
The signal/noise power measurement, which is essential for algorithms, and the implementation of AGC, can be performed with the assistance of a high-resolution ADC. 
For MIMO systems, applying mixed ADC architectures \cite{TJiang-2016TWC}, namely, equipping few high-resolution ADCs for improving detection and estimation accuracy, can also help signal/noise power measurement. 
Moreover, in the presence of coarse quantization, conventional correlation-based synchronization remains valid.

The rest of the paper is organized as follows.
Section II models the Q-OFDM system, and highlights the inference problems and the system implementation issues considered in this study.
Section III proposes algorithms based on GTurbo framework.
Section IV evaluates and compares the performances of the proposed algorithms through the numerical simulation results.
Section V presents the proposed receiver architecture, the construction of the prototyping system and the OTA experiment results.
Section VI discusses our ideas of future extensions.
Finally, Section VII concludes this paper.

\emph{Notations:} This paper uses lowercase and uppercase boldface letters to represent vectors and matrices, respectively.
For vector $\mathbf{a}$, $a_j$ denotes the $j$-th entry of $\mathbf{a}$, $\|\mathbf{a}\|$ denotes the $l_2$ norm of the vector $\mathbf{a}$, and the operator $\mathrm{diag}(\mathbf{a})$ denotes the diagonal matrix with diagonal elements as the entries of $\mathbf{a}$.
For matrix $\mathbf{A}$, the operator $\mathrm{Diag}(\mathbf{A})$ denotes the vector composed of the diagonal elements of matrix $\mathbf{A}$.
Moreover, the real and imaginary parts of a complex scalar $a$ are represented by $a^R$ and $a^I$, respectively.
For random vectors $\mathbf{a}$ and $\mathbf{b}$, the notation $\mathbf{R}_{\mathbf{a}\mathbf{b}}=\mathrm{E}\{\mathbf{a}\mathbf{b}^{H}\}$ represents their cross-correlation matrix.
The distribution of a proper complex Gaussian random variable $z$ with mean $\mu$ and variance $\nu$ is denoted by $\mathcal{CN}(z;\mu,\nu)$.
Similarly, $\mathcal{N}(z;\mu,\nu)$ denotes the probability density function of a real Gaussian random variable $z$ with mean $\mu$ and variance $\nu$.
The probability density function and cumulative distribution function for the standard Gaussian distribution are denoted by $\phi(z)=\frac{1}{\sqrt{2\pi}}e^{-\frac{z^2}{2}}$ and  $\Phi(z)=\frac{1}{\sqrt{2\pi}}\int_{-\infty}^{z}e^{-\frac{t^2}{2}}\mathrm{d}t$ respectively, and the Q function is defined as $Q(z)=1-\Phi(z)$.

\section{System Model}

We consider an OFDM system that comprises $N$ orthogonal subcarriers, where $N_d$ of them  corresponding to frequency $-\frac{N_d}{2}\frac{F_s}{N},\ldots,-\frac{F_s}{N},\frac{F_s}{N},\ldots,\frac{N_d}{2}\frac{F_s}{N}$ ($F_s$ denotes the sampling frequency) are assigned to carry data symbols.
Meanwhile, the rest of the $N-N_d$ subcarriers are reserved as guard bands, which are padded by  zero signals to reduce signal leakage into the adjacent frequency bands.
Hereinafter, we use $\mathcal{X}=\{1,2,\cdots,N\}$ to denote the index set of all subcarriers, and $\mathcal{X}_d\subseteq\mathcal{X}$ to denote the index subset of the $N_d$ subcarriers used for data transmission.
In addition, the subscript ``$_d$'' for the $N$-dimension vector and a matrix with $N$ rows (e.g., $\mathbf{a}_d$ and $\mathbf{A}_d$) represent the $N_d$-dimension subvector that comprises the elements of $\mathbf{a}$ with index $\mathcal{X}_d$ and the submatrix that comprises the rows of $\mathbf{A}$ with index $\mathcal{X}_d$, respectively.

\begin{figure}[!t]
	\centering
	\includegraphics[scale=0.4]{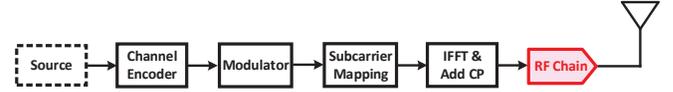}
	\caption{Block diagram of an OFDM transmitter.\label{F21}}
\end{figure}

The signaling process at the OFDM transmitter is illustrated in Fig. \ref{F21}.\footnote{We only specify here the signaling process corresponding to the OFDM symbols used for data transmission. For OFDM symbols for other purposes, specific baseband symbols (e.g., pilot symbols and synchronization sequences) are directly assigned into their corresponding subcarriers without the need of channel encoding and constellation mapping.}
The information bits to be transmitted, which were generated by the data source, are first encoded using a channel code technique, such as the turbo code.
Then, the modulator maps the coded bits into the corresponding constellation symbols based on the selected modulation method, such as $M$-ary quadrature amplitude modulation ($M$-QAM), and each modulated symbol is assigned into one of the $N_d$ subcarriers of the OFDM symbol.
We denote the vector of the frequency-domain constellation symbols as ${\mathbf{s}\in\mathbb{C}^{N\times 1}}$, where ${s_j=0}$ for ${j\in\mathcal{X}\backslash\mathcal{X}_d}$.
Subsequently, the frequency-domain OFDM symbol $\mathbf{s}$ is transformed into the time-domain using $N$-point IDFT.
The CP with the length $L_{\rm{cp}}$ greater than the length of channel impulse response is appended at the beginning of each OFDM symbol.
This contributes to convert inter-symbol interference (ISI) channel caused by the multipath effect into a bank of flat-fading subchannels. 
The obtained sequence of the time-domain baseband samples is finally up-converted by the RF chain and transmitted over the wireless channel.

At the receiver, the RF signal received by the antenna is first down-converted into the analog baseband.
Then the analog baseband signal is sampled with frequency $F_s$ and quantized using an ultra-low resolution ADC.
After CP removal and under the assumption of an ideal synchronization, the coarsely quantized time-domain samples of an OFDM symbol can be expressed as
\begin{equation}\label{DET_Prob_Form}
	 \mathbf{q}_D=\mathcal{Q}_c{\left[\sqrt{g_{\rm{AGC}}}\left(\mathbf{F}^H\mathrm{diag}({\mathbf{h}})\mathbf{s}+\mathbf{n}\right)\right]},
\end{equation}
where $\mathbf{F}$ denotes the normalized DFT matrix with $(m,n)$-th entry being $\frac{1}{{\sqrt N }} e^{-{2\pi j(n - 1)(m - 1)}/N}$;
$\mathbf{n}\in\mathbb{C}^{N\times 1}$ denotes the AWGN vector with distribution $\mathcal{CN}(0,\sigma^2\mathbf{I})$;
$\mathbf{h}$ is the frequency-domain channel response vector, which is obtained by performing DFT on the tapped delay line representation of the ISI channel denoted by $\mathbf{h}_{{\rm t}} = [h_{{\rm t},1}, \, h_{{\rm t},2}, \cdots, h_{{\rm t},N}]^T$ with ${h_{{\rm t},j}=0}$ for ${(L+1)\leq j\leq N}$ and $L$ being the maximum number of channel taps;
$g_{\rm{AGC}}$ is the gain of the AGC unit used to adjust the amplitude of the ADC input within a certain range;
$\mathcal{Q}_c(\cdot)$ denotes the elementwise mapping function of the complex-valued quantizer, which comprises two parallel real-valued $B$-bit quantizers $\mathcal{Q}(\cdot)$ that independently quantize the real and imaginary parts of each analog input sample.
For example, the quantization of a scalar analog sample $y$ will be
\[\mathcal{Q}_c(y)=\mathcal{Q}(y^R)+j\mathcal{Q}(y^I),\]
where quantizer $\mathcal{Q}(\cdot)$ outputs the discrete level $c_b$ when the real-valued input $y^R$ or $y^I$ is within the interval $(r_{b-1},r_{b}]$, with $-\infty=r_{0}<r_1<\cdots<r_{2^{B-1}}<r_{2^B}=\infty$ being the thresholds of the quantizer. 
Notably, \eqref{DET_Prob_Form} can also be used as an equivalent input-output relation of OFDM systems with multi-antenna and analog beamforming when tackling data detection and channel estimation. 
Therefore, $\mathbf{h}$ denotes the beamformed channel response in frequency domain, like the way in \cite{Yin-13GLCOM}.

We design the baseband processing architecture at the receiver to recover the original information bits from each coarsely quantized received OFDM symbol $\mathbf{q}_D$ with error rate given the channel estimate.
Hereinafter we use the notation $\bar{\mathbf{h}}$ to denote the product of the estimate of $\mathbf{h}$ and $\sqrt{g_{AGC}}$.
For example, the conventional OFDM receiver first transforms the $\mathbf{q}_D$ into the frequency-domain, which yields $\tilde{\mathbf{q}}=\mathbf{F}\mathbf{q}_D$, followed by the one-tap equalizer, which computes the estimate of $s_j$ for $j\in\mathcal{X}_d$ by ${{\tilde q}_j}/{{\bar h}_j} $.
However, this conventional strategy fails to work well in Q-OFDM systems because the orthogonality among subcarriers is destroyed.
The iterative procedure for the MMSE estimation of $\mathbf{s}$ was proposed in \cite{Wang-2017JSAC} under the assumption of a perfect CSI, noise power, and synchronization.
Determining how to realize those assumptions comprises the content of this paper. In particular,

\emph{1) Channel Estimation}:
To obtain CSI, we periodically assign an OFDM symbol that contains pilot symbols at the subcarriers with index $\mathcal{X}_d$.
The vector of the frequency-domain pilot symbols is denoted by ${\mathbf{p}\in\mathbb{C}^{N}}$, where ${p_j=0}$ for ${j\in\mathcal{X}\backslash\mathcal{X}_d}$.
The quantized received signal vector for each pilot-transmitting OFDM symbol can be expressed as
\begin{equation}\label{CE_Prob_Form}
	\mathbf{q}_P=\mathcal{Q}_c{\left[\sqrt{g_{\rm{AGC}}}\left(\mathbf{F}^H\mathrm{diag}({\mathbf{p}})\mathbf{h}+\mathbf{n}\right)\right]}.
\end{equation}
For notation simplification, we define $\bar{\mathbf{p}}=\sqrt{g_{\rm{AGC}}}\,\mathbf{p}$.
The conventional OFDM channel estimator performs least-squares (LS) channel estimation to first yield ${{{\bf{F}}{{\bf{q}}_P}} \mathord{\left/
			{\vphantom {{{\bf{F}}{{\bf{q}}_P}} {{\bf{\bar p}}}}} \right.
			\kern-\nulldelimiterspace} {{\bf{\bar p}}}}$, and then left-multiplies it by a certain weight matrix.
This traditional approach relies heavily on orthogonality among subcarriers, which is already destroyed by the low-resolution ADC.
This fact and the success of GTurbo in Q-OFDM detection have motivated us to develop a specialized channel estimator for Q-OFDM based on this advanced signal processing technique.

\emph{2) Implementation Issues}: A number of issues should be considered in implementing the prototyping system, including adjusting the amplitude of the baseband signal within a certain range (i.e., AGC), measuring signal/noise power, and identifying the starting point of each frame, and subsequently, each OFDM symbol (i.e., synchronization).
In conventional systems with high-resolution ADCs, the first two issues can be addressed by using a power accumulator, which is applied by taking the long-term time average of the amplitude of the corresponding received baseband samples. Meanwhile, synchronization is achieved through transmitting a pre-defined complex primary synchronization sequence (PSS) in each radio frame, computing the correlation between the received signal and the reference sequence, and finding the local maximum correlation to identify the starting point of each frame.
We should clarify if these conventional solutions are still valid for the Q-OFDM receiver.

\section{GTurbo-based Algorithms}

The GTurbo signal recovery principle was proposed in \cite{CKWen-2016ISIT} for a quantized compressed sensing problem with an orthogonal sensing matrix.
It consists of two concatenated modules. The first module produces a coarse estimate from the received quantized signal; the second module refines the estimate by considering the prior of the signal to be estimated.
Each module computes its extrinsic information to be the input of the other module.
The two modules are executed iteratively until convergence is achieved.
This is why we use ``turbo`` to name it.
GTurbo is a powerful tool for an inference problem formulated as a \emph{generalized linear model} with an orthogonal linear transformation matrix.
The Q-OFDM detection problem formulated in \eqref{DET_Prob_Form} belongs to this class of problem with a DFT matrix.
However, the original GTurbo is only applicable to cases where the priori of the estimated signal is available.
In this section, we extend the original GTurbo to the problem without a known priori and propose an efficient channel estimator for a Q-OFDM system.

\subsection{Channel Estimator}

The purpose of a channel estimator is to estimate the frequency-domain channel parameters of subcarriers dedicated for data transmission, namely $\mathbf{h}_d$, based on the given $ \mathbf{q}_P$ and $\bar{\mathbf{p}}$.
The estimated channel parameter $\bar{\mathbf{h}}_d$ is subsequently utilized as the CSI for data detection to the remainder of the OFDM symbols in the current slot.
Before formally introducing the proposed algorithm, we define the following two auxiliary vectors to facilitate our subsequent discussion: 
\begin{equation} \label{aux_vec}
\mathbf{x}:=\mathrm{diag}(\bar{\mathbf{p}})\mathbf{h},~~~ \mathbf{z}:=\mathbf{F}^H\mathbf{x},
\end{equation}
where $\mathbf{x}$ and $\mathbf{z}$ represent the noiseless signals received in the frequency and time domains, respectively. 

\begin{figure}[!t]
	\centering
	\includegraphics[scale=0.45]{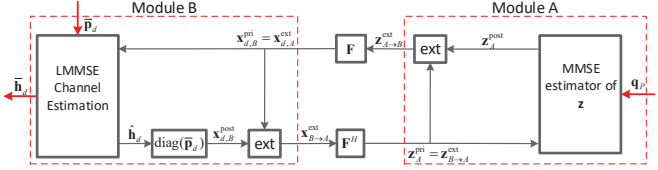}
	\caption{ Block diagram of the GTurbo-LMMSE channel estimator. The ``ext'' block represents the extrinsic information computation. The block of a certain matrix represents the left-multiplying operation of the input vector by the matrix in the block.\label{F3}}
\end{figure}

The channel estimation problem \eqref{CE_Prob_Form} has been formulated as a generalized linear model of the frequency-domain channel vector $\mathbf{h}$,
suggesting that the GTurbo principle \cite{CKWen-2016ISIT} can be applied for estimating $\mathbf{h}_d$.
The proposed channel estimation technique is illustrated in Fig. \ref{F3}.
The GTurbo-based algorithm comprises two processing modules.
Module A is used to directly produce a coarse estimate of $\mathbf{x}$ from the quantized measurement $\mathbf{q}_P$, which  is equivalent to producing a coarse estimate of $\mathbf{h}$ following the linear transform in \eqref{aux_vec}.
Module B is used to produce a refinement of the estimate of $\mathbf{h}_d$.
The extrinsic mean and variance of each module are passed to the other module as input.
In each iteration, Modules A and B are implemented successively to refine the estimate of $\mathbf{h}_d$ until convergence is achieved. 

The proposed channel estimation algorithm is presented in Algorithm \ref{A1}. We use $t$ to denote the iteration index.
We then provide several intuitive explanations to allow readers to improve their understanding of the proposed channel estimator.
In Module A, \eqref{post_mean_z} and \eqref{post_var_z} compute the MMSE estimate of $z_j$ under the given quantized observation $q_j$, and its corresponding MSE, where $\mathrm{E}\left[z^R_j\mid q^R_{j,P}\right]$ and $\mathrm{var}\left[z^R_j\mid q^R_{j,P}\right]$ denote the posteriori expectation and variance of $z^R_j$ given $q^R_{j,P}$ under the assumption of $\mathrm{P}(z^R_j)=\mathcal{N}(z^R_j;z_{j,A}^{{\rm pri},R}(t),\frac{1}{2}v_A^{{\rm pri}}(t))$. Their explicit expressions can be derived from \cite{CKWen-2016TSP} as
\begin{subequations}
	\begin{align}
	&\mathrm{E}\left[z^R_j\mid q^R_{j,P}\right]=z_{j,A}^{{\rm pri},R}+\frac{v_A^{{\rm pri}}(t)}{\sqrt{2(v_A^{{\rm pri}}(t)+\overline {{\sigma ^2}} )}}\left(\frac{\phi(\eta_1)-\phi(\eta_2)}{\Phi(\eta_1)-\Phi(\eta_2)}\right),\label{eIII_01}\displaybreak[0]\\
	&\begin{aligned}
	&\mathrm{var}{\left[z^R_j\mid q^R_{j,P}\right]}=\frac{v_A^{{\rm pri}}(t)}{2}-\frac{(v_A^{{\rm pri}}(t))^2}{2(v_A^{{\rm pri}}(t)+\overline {{\sigma ^2}} )}\times\\
	&~~~~~~{\left[\left(\frac{\phi(\eta_1)-\phi(\eta_2)}{\Phi(\eta_1)-\Phi(\eta_2)}\right)^2+\frac{\eta_1\phi(\eta_1)-\eta_2\phi(\eta_2)}{\Phi(\eta_1)-\Phi(\eta_2)}\right]},
	\end{aligned} \label{eIII_02}
	\end{align}
\end{subequations}
where $\overline {{\sigma ^2}}  = {g_{{\rm{AGC}}}} \widehat{\sigma^2}$ with $\widehat{\sigma^2}$ being the estimated noise
power, and
\begin{equation}\label{eIII_03}
\eta_1=\frac{z_{j,A}^{{\rm pri},R}(t)-l(q^R_{j,P})}{\sqrt{(v_A^{{\rm pri}}(t)+\overline {{\sigma ^2}} )/2}},~~\eta_2=\frac{z_{j,A}^{{\rm pri},R}(t)-u(q^R_{j,P})}{\sqrt{(v_A^{{\rm pri}}(t)+\overline {{\sigma ^2}} )/2}}
\end{equation}
where $l(q^R_{j,P})$ and $u(q^R_{j,P})$ are the lower and upper bounds that correspond to the quantizer output value $q^R_{j,P}$.
Moreover, $\mathrm{E}{\left[z^I_j\mid q^I_{j,P}\right]}$ and $\mathrm{var}{\left[z^I_j\mid  q^I_{j,P}\right]}$ can be similarly computed by simply replacing $z_{j,A}^{{\rm pri},R}(t)$ with $z_{j,A}^{{\rm pri},I}(t)$ when computing $\eta_1$ and $\eta_2$ in \eqref{eIII_03}.
Subsequently, the extrinsic information of Module A can be computed using \eqref{post_var_A}--\eqref{ext_mean_A2}. In particular, \eqref{post_var_A} and \eqref{ext_var_A} compute the extrinsic variance of Module A, whereas \eqref{ext_mean_A1} and \eqref{ext_mean_A2} compute the extrinsic mean of Module A, which is then used as an input of Module B.

\begin{algorithm}[!ht]
	\caption{GTurbo-LMMSE Channel Estimator\label{A1}}
	\small
	\KwData{$\mathbf{z}_A^{{\rm pri}}(1)=\mathbf{0}_{N\times 1}$, $v_A^{{\rm pri}}(1)=1-g_{AGC}\widehat {{\sigma ^2}}$, $\mathbf{x}_B^{{\rm pri}}=\mathbf{0}_{N\times 1}$, $\mathbf{x}_{B}^{\rm post}=\mathbf{0}_{N\times 1}$\;}
	\For{$t=1:T_{\mathrm{max}}$}{
		\textbf{Module A:}\\
		(1) Compute the a posteriori mean/variance of $\mathbf{z}$:
		\begin{subequations}
			\begin{align}
			&z_{j,A}^{\rm post}=\mathrm{E}\left[z^R_j\mid  q^R_{j,P}\right]+j\mathrm{E}\left[z^I_j\mid  q^I_{j,P}\right],\label{post_mean_z}\\
			&v_{j,A}^{\rm post}=\mathrm{var}\left[z^R_j\mid  q^R_{j,P}\right]+\mathrm{var}\left[z^I_j\mid q^I_{j,P}\right],\label{post_var_z}
			\end{align}
		\end{subequations}
		(2) Compute the extrinsic information of Module A:
		\begin{subequations} \label{ext_mean_var_A}
			\begin{align}
			&v_{A}^{\rm post}=\frac{1}{N}\sum\limits_{j=1}^{N}v_{j,A}^{\rm post},\label{post_var_A}\\
			&v_B^{{\rm pri}}=v_{A\to B}^{\rm ext}=\left(\frac{1}{v_{A}^{\rm post}}-\frac{1}{ v_A^{{\rm pri}} (t)}\right)^{-1},\label{ext_var_A}\\
			&\mathbf{x}_{A}^{\rm ext}=v_{A\to B}^{\rm ext}\left(\frac{\mathbf{F}\mathbf{z}_A^{\rm post}}{v_{A}^{\rm post}}-\frac{\mathbf{F} \mathbf{z}_A^{{\rm pri}}(t)}{v_A^{{\rm pri}}(t)}\right),\label{ext_mean_A1}\\
			&\mathbf{x}_{d,B}^{{\rm pri}}=\mathbf{x}_{d,A\to B}^{\rm ext},\label{ext_mean_A2}
			\end{align}
		\end{subequations}
		\textbf{Module B:}\\
		(3) LMMSE channel estimation:
		\begin{equation}
		\hat{\mathbf{h}}_d=\mathbf{W}_{\mathrm{LMMSE}}\mathrm{diag}^{-1}(\bar{\mathbf{p}}_d)\mathbf{x}_{d,B}^{{\rm pri}},\label{LMMSE_denoiser}
		\end{equation}
		(4) Compute the extrinsic information of Module B:
		\begin{subequations} \label{eq:e0530}
			\begin{align}
			&\mathbf{x}_{d,B}^{\rm post} = \mathrm{diag}(\bar{\mathbf{p}}_d)\hat{\mathbf{h}}_d,\label{post_mean_A}\\
			&\mathbf{x}_A^{{\rm pri}}=\mathbf{x}_{B\to A}^{\rm ext}=c\left(\mathbf{x}_B^{\rm post}-\alpha\mathbf{x}_B^{{\rm pri}}\right), \label{ext_mean_B}\\
			&\mathbf{z}_A^{{\rm pri}}(t+1)=\mathbf{F}^H\mathbf{x}_A^{{\rm pri}},\\
			&v_A^{{\rm pri}}(t+1)=v_{B\to A}^{\rm ext}=\frac{\left|\left|\mathbf{x}_{d,B\to A}^{\rm ext}-\mathbf{x}_{d,B}^{\rm post}\right|\right|^2}{N_d}\label{ext_var_B}.	
			\end{align}
		\end{subequations}
	}
\end{algorithm}

We then consider the operations of Module B for extracting the desired channel parameters.
The input of Module B $\mathbf{x}_B^{{\rm pri}}$ can be approximately represented by the AWGN observation of $\mathbf{x}=\mathrm{diag}(\bar{\mathbf{p}})\mathbf{h}$ as follows:
\begin{equation}\label{appro_AWGN}
\mathbf{x}_B^{{\rm pri}}=\mathrm{diag}(\bar{\mathbf{p}})\mathbf{h}+\bm{\omega}_B,
\end{equation}
where $\bm{\omega}_B\sim\mathcal{CN}(\mathbf{0},v_B^{{\rm pri}}\bf{I})$.
To be continuous, the following challenges should be addressed:
\begin{itemize}
	\item The MMSE estimate of $\mathbf{h}_d$ from \eqref{appro_AWGN} as done in the original GTurbo framework requires the prior distribution of $\mathbf{h}_d$, which is typically unavailable in practical systems.
    Hence the operation of Module B needs to be re-designed. 
    
	\item The derivation of the extrinsic mean and variance that correspond to the preceding operations guarantees the convergence of the proposed algorithm.
\end{itemize}

The first issue can resort to the well-developed ``denoiser'' in the image processing field.
Various types of denoisers, such as SURE-LET \cite{Luisier-2007TIP}, BM3D \cite{Egiazarian-2006SPIE}, dictionary learning \cite{Aharon-2006TIP}, and SVT \cite{Candes-2010SIAMJO}, are available.
However, these denoisers are specialized for i.i.d. sparse signal, low-rank matrix, or image.
In our earlier trial computer simulations, the mechanical application of these denoisers cannot achieve satisfactory performance in estimating $\mathbf{h}_d$ with strongly correlated entries.
Therefore, we must select a new type of denoiser for this channel estimation problem.

Extracting $\mathbf{h}_d$ from its approximate AWGN observation \eqref{appro_AWGN} is similar to the channel estimation problem in conventional full-resolution OFDM systems.
The LMMSE denoiser for extracting $\mathbf{h}_d$ from $\mathbf{x}_B^{{\rm pri}}$ is constructed using \eqref{LMMSE_denoiser}, where $\mathbf{W}_{\mathrm{LMMSE}}$ is the corresponding linear weight matrix with the form
\begin{equation}\label{LMMSE_Mat}
	\mathbf{W}_{\mathrm{LMMSE}}=\mathbf{R}_{\mathbf{h}\mathbf{h}_d}{\left(\mathbf{R}_{\mathbf{h}_d\mathbf{h}_d}+\gamma^2\mathbf{I}\right)}^{-1},
\end{equation}
where $\mathbf{R}_{\mathbf{h}\mathbf{h}_d} = \mathrm{E}[ \mathbf{h}\mathbf{h}^{H}_d ]$ and $\mathbf{R}_{\mathbf{h}_d\mathbf{h}_d} = \mathrm{E}[ \mathbf{h}_d\mathbf{h}^{H}_d ]$ are the channel cross-correlation and auto-correlation matrices, respectively; and $\gamma^2$ is the variance of additive noise in \eqref{appro_AWGN}.
However, to implement a strict-sense LMMSE denoiser, $\mathbf{R}_{\mathbf{h}\mathbf{h}_d}$ and $\mathbf{R}_{\mathbf{h}_d\mathbf{h}_d}$ should be known.
Therefore, we assume the time-domain profile of the channel to be a uniformly tapped delay line, with the maximum number of channel taps as $\hat{L}$.
Given this assumption, the $(m,n)$-th entry of $\mathbf{R}_{\mathbf{h}_d\mathbf{h}_d}$ can be computed using \cite[Appendix A]{Edfors1998TCOM}
\[{\left[ {{{\bf{R}}_{{{\bf{h}}_d}{{\bf{h}}_d}}}} \right]_{(m,n)}} = \left\{ {\begin{array}{*{20}{l}}
	{1,}&{{{\mathcal{X}}_d}\left( m \right){\mathcal{ = }}{{\mathcal{X}}_d}(n)}\\
	{\frac{{1 - {e^{ - j2\pi \hat{L} \frac{{{{\mathcal{X}}_d}\left( m \right) - {{\mathcal{X}}_d}(n)}}{N}}}}}{{j2\pi \hat{L} \frac{{{{\mathcal{X}}_d}\left( m \right) - {{\mathcal{X}}_d}(n)}}{N}}},}&{{{\mathcal{X}}_d}\left( m \right) \ne {{\mathcal{X}}_d}(n)}
	\end{array}} \right.\]
where ${{\mathcal{X}}_d}(n)$ is the $n$-th element of set ${{\mathcal{X}}_d}$.
And entries of $\mathbf{R}_{\mathbf{h}\mathbf{h}_d}$ can be computed similarly.
In addition, we let $\gamma^2$ be an extremely small number compared with the variance of additive noise.
The preceding assumptions have been proven by \cite{Li-2000TVT} to perform close to the strict-sense LMMSE estimation. 
Furthermore, the reliability and hardware efficiency of the approximate LMMSE channel estimator are confirmed by our previously developed massive MIMO-OFDM prototyping system \cite{Gao-2018ACCESS}.

The remaining task is to derive the extrinsic mean and variance of Module B that are associated with the approximate LMMSE denoiser.
Following the divergence-free property of the series of algorithms extended from the turbo principle \cite{JMa-2016Access},
we first construct the extrinsic mean $\mathbf{x}_{B\to A}^{\rm ext}$ as the linear combination of $\mathbf{x}_B^{\rm post}$ and $\mathbf{x}_B^{{\rm pri}}$ with the form of \eqref{ext_mean_B},
and then determine coefficients $\alpha$ and $c$ based on the following two conditions:
\begin{itemize}
\item The errors of the input vector and extrinsic mean of Module B are orthogonal, namely,
\begin{equation}\label{div_free}
\mathrm{E}[(\mathbf{x}_{B}^{\rm pri}-\mathbf{x})^{H}(\mathbf{x}_{B\to A}^{\rm ext}-\mathbf{x})]=0.
\end{equation}

\item The error of the extrinsic mean of Module B, i.e., $\mathrm{E}\left[\left\|\mathbf{x}_{B\to A}^{\rm ext}-\mathbf{x}\right\|^2\right]$, is minimized.
\end{itemize}

Coefficients $\alpha$ and $c$ that satisfy the aforementioned conditions can be obtained by following the derivations in \cite{JMa-2017Access}, which is given by
\begin{subequations}\label{ext_inf}
	\begin{align}
	&\alpha=\frac{\mathbf{1}_{1\times N_d }\mathrm{diag}^{-1}(\bar{\mathbf{p}}_d)\mathrm{Diag}(\mathbf{W}_{d,\mathrm{LMMSE}})}{N_d},\\
	&c=\frac{(\mathbf{x}_{B}^{\rm pri})^{H}(\mathbf{x}_{B}^{\rm post}-\alpha\mathbf{x}_{B}^{\rm pri})}{\|\mathbf{x}_{B}^{\rm post}-\alpha\mathbf{x}_{B}^{\rm pri}\|^2},
	\end{align}
\end{subequations}
where $\mathbf{1}_{1\times N_d }$ denotes the $N_d$-dimension row vector whose elements are all one.
In addition, we can approximate the extrinsic variance $v_{B\to A}^{\rm ext}=\frac{1}{N_d}||\mathbf{x}_{d,B\to A}^{\rm ext}-\mathbf{x}_d||^2$ by relaxing the exact $\mathbf{x}_d$ as its estimate $\mathbf{x}_{d,B}^{\rm post}$.

Notably, nearly all amplitude information  is lost irretrievably under 1-bit quantization.
We must recover amplitude information, particularly the average channel gain $P_h=\frac{1}{N_d}\sum_{j\in{\mathcal{X}}_d}|h_j|^2$, via additional means to overcome this intrinsic defect of the 1-bit ADC and further improve  performance. We refine the channel estimate $\hat{\mathbf{h}}_d$ with the additional amplitude information by operating the channel normalization expressed by
\begin{equation}\label{Channel_norm}
\tilde{\mathbf{h}}_d = \sqrt{\hat{P}_h}\frac{{{{{\bf{\hat h}}}_d}}}{{\| {{{{\bf{\hat h}}}_d}} \|}},
\end{equation}
which is the final output of the channel estimator.
In our system, the estimated $P_h$ can be obtained from the average power of the received signal $P_r$, which is defined by
\begin{equation}\label{est_norm}
\hat{P}_h = {\left(P_r - \widehat {{\sigma ^2}}\right)}\times\frac{N}{N_d}.
\end{equation}
We will specify how to measure $P_r$ in Section V.
Then, $\bar{\mathbf{h}}_d=\sqrt{g_{\rm{AGC}}}\tilde{\mathbf{h}}_d$ will be used as the CSI for the data detector after obtaining the normalized channel estimate  $\tilde{\mathbf{h}}_d$.
Channel normalization is unnecessary when ADC precision is more than 1 bit because the majority of amplitude information can be recovered using the proposed GTurbo-LMMSE channel estimator.
This result can be verified by the numerical results presented in Section IV.
Let $\bar{\mathbf{h}}_d=\sqrt{g_{\rm{AGC}}}\hat{\mathbf{h}}_d$ be the CSI for the data detector under the case without channel normalization.

\subsection{Data Detection and Channel Decoding}
The data detector is used to compute the MMSE estimate of $\mathbf{s}$ and its corresponding MSE based on the quantized received signal $\mathbf{q}_D$ modeled by \eqref{DET_Prob_Form}.
The GTurbo-based detector proposed in \cite[Algorithm 1]{Wang-2017JSAC} can be directly used in data detection with a slight modification by replacing the known CSI with the estimated channel parameters $\bar{\mathbf{h}}$.
In this subsection, we briefly describe the processing procedure for the data detector.
For the details of the algorithm, refer to \cite{Wang-2017JSAC}.
\begin{figure}[!t]
	\centering
	\includegraphics[scale=0.45]{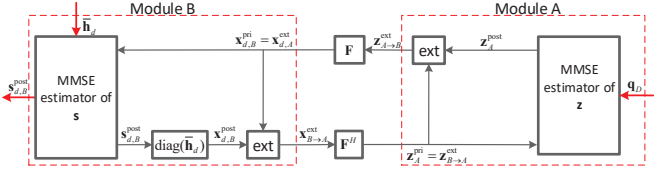}
	\caption{ Block diagram of the GTurbo-based data detector with an estimated CSI $\bar{\mathbf{h}}$. The ``ext'' block represents the extrinsic information computation. 
		The block of a certain matrix represents the left-multiplying operation of the input vector by the matrix in the block.\label{F4}}
\end{figure}

We also define two auxiliary vectors similar to \eqref{aux_vec} as follows:
\[\mathbf{x}:=\mathrm{diag}(\bar{\mathbf{h}})\mathbf{s},~~~ \mathbf{z}:=\mathbf{F}^H\mathbf{x}.\]
The computation procedures are demonstrated in Algorithm \ref{A2}, and the corresponding flow diagram of this algorithm is presented in Fig. \ref{F4}.
the two modules of this detector are as follows.
Module A calculates the coarse estimate of $\mathbf{x}$ without considering the prior distribution of $\mathbf{s}$.
Module B refines the estimate by considering the prior distribution of $\mathbf{s}$.
In each iteration, Modules A and B are performed successively to refine the estimate of $\mathbf{s}$ until convergence is achieved.

In Module A, \eqref{eA1_01} and \eqref{eA1_02} compute the posteriori mean and variance of $z_j$ with explicit expressions provided by  \eqref{eIII_01} and \eqref{eIII_02}, respectively.
The extrinsic variance of Module A can be computed using \eqref{eA1_04} and \eqref{eA1_05}, whereas its extrinsic mean can be computed using \eqref{eA1_06} and \eqref{eA1_09}, respectively.
In Module B, we compute the estimate of $\mathbf{s}$ and its MSE for $j\in\mathcal{X}_d$ in \eqref{eA1_07} and \eqref{eA1_08}. Their explicit expressions are given by
\begin{subequations}
	\begin{align}
	&s_{j,B}^{\rm post} = \frac{{\sum\limits_{ s\in\mathcal{S} } {s\, \mathcal{CN}\left(s;\frac{{x_{j,B}^{{\rm pri}}}}{{{\bar{h}_j}}},\frac{{v_B^{{\rm pri}}}}{{{{\left| {{\bar{h}_j}} \right|}^2}}}\right)} }}{{ \sum\limits_{ s\in\mathcal{S} } { \mathcal{CN}\left(s;\frac{{x_{j,B}^{{\rm pri}}}}{{{\bar{h}_j}}},\frac{{v_B^{{\rm pri}}}}{{{{\left| {{\bar{h}_j}} \right|}^2}}}\right)} }},\displaybreak[0]\label{eIII_05}\\
	&v_{j,B}^{\rm post} = \frac{{\sum\limits_{ s\in\mathcal{S} } { {{\left| {s} \right|}^2}\mathcal{CN}\left(s;\frac{{x_{j,B}^{{\rm pri}}}}{{{\bar{h}_j}}},\frac{{v_B^{{\rm pri}}}}{{{{\left| {{\bar{h}_j}} \right|}^2}}}\right)} }}{{ \sum\limits_{ s\in\mathcal{S} } { \mathcal{CN}\left(s;\frac{{x_{j,B}^{{\rm pri}}}}{{{\bar{h}_j}}},\frac{{v_B^{{\rm pri}}}}{{{{\left| {{\bar{h}_j}} \right|}^2}}}\right)} }} - {\left| {s_{j,B}^{\rm post}} \right|^2}.\label{eIII_06}
	\end{align}
\end{subequations}
where $\mathcal{S}$ denotes the set of constellation points.
Subsequently, we compute the extrinsic variance of Module B using \eqref{e0530_07b} and \eqref{e0411_08a} and its extrinsic mean using \eqref{e0530_07a} and \eqref{e0411_08b}.
\eqref{e0411_08b} is identical to \eqref{ext_mean_B} by letting $c=\frac{v_{B\to A}^{\rm ext}}{v_{B}^{\rm post}}$ and $\alpha=\frac{v_{B}^{\rm post}}{v_B^{{\rm pri}}}$, which also satisfy conditions (a) and (b).

In the practical applications, we can set an additional termination criterion, namely, the iteration should be terminated when $v_B^{{\rm pri}}$ or $v_A^{{\rm pri}}$ is less than the predefined thresholds.
This modification is reasonable because that the excessively small $v_B^{{\rm pri}}$ and $v_A^{{\rm pri}}$, along with errors in the estimated channel responses, possibly trap the estimates \eqref{eA1_01} and \eqref{eA1_07} into wrong values, which deteriorates detection performance.
Finally, $\mathbf{x}_{d,B}^{{\rm pri}}$ and $v_B^{\rm pri}$ are used for the computation of the log-likelihood ratio (LLR) that corresponds to each coded bit, which is used as the input of the channel decoder.
We denote the $i$-th bit associated with the modulated symbol $s_j$ by $b_{ji}$. The LLR of $b_{ji}$ can be computed by
\begin{equation}\label{Cal_LLR}
{\rm{LLR}}\left( {{b_{ji}}} \right) = \frac{1}{{\beta_j }}\left( {{{\left| {\frac{{x_{j,B}^{{\rm pri}}}}{{{\bar{h}_j}}}- s_{j,i}^ - } \right|}^2} - {{\left| {\frac{{x_{j,B}^{{\rm pri}}}}{{{\bar{h}_j}}} - s_{j,i}^ + } \right|}^2}} \right),
\end{equation}
where $\beta_j  = v_B^{\rm pri}\left/{\left| {{\bar{h}_j}} \right|^2}\right.$, $s_{j,i}^ +$ and $s_{j,i}^ -$ are defined as follows:
\begin{align}\notag
&s_{j,i}^{-} = \mathop{{\rm argmin}}\limits_{s\in\mathcal{S}_{i}^{-}} \left|s-\frac{{x_{j,B}^{{\rm pri}}}}{{{\bar{h}_j}}}\right|^2,\notag\displaybreak[0]\\
&s_{j,i}^{+} = \mathop{{\rm argmin}}\limits_{s\in\mathcal{S}_{i}^{+}} \left|s-\frac{{x_{j,B}^{{\rm pri}}}}{{{\bar{h}_j}}}\right|^2,\notag
\end{align}
where $\mathcal{S}_{i}^{+}$ and $\mathcal{S}_{i}^{-}$ denote the subsets of the constellation symbols of the adopted modulation scheme with $i$-th bit being 1 and 0, respectively.

\begin{algorithm}[!ht]
	\caption{GTurbo-based Data Detector\label{A2}}
	\small
	\KwData{$\mathbf{z}_A^{{\rm pri}}(1)=\mathbf{0}_{N\times 1}$, $v_A^{{\rm pri}}(1)=\frac{1}{N}\sum_{j\in{\mathcal{X}}_d}|\bar{h}_j|^2$, $\mathbf{x}_B^{{\rm pri}}=\mathbf{0}_{N\times 1}$, $\mathbf{x}_{B}^{\rm post}=\mathbf{0}_{N\times 1}$\;}
	\For{$t=1:T_{\mathrm{max}}$}{
		\textbf{Module A:}\\
		(1) Compute the posteriori mean/variance of $\mathbf{z}$:
		\begin{subequations} \label{eA1}
			\begin{align}
			&z_{j,A}^{\rm post}=\mathrm{E}\left[z^R_j\mid  q^R_{j,D}\right]+j\mathrm{E}\left[z^I_j\mid q^I_{j,D}\right],\label{eA1_01}\\
			&v_{j,A}^{\rm post}=\mathrm{var}\left[z^R_j\mid  q^R_{j,D}\right]+\mathrm{var}\left[z^I_j\mid  q^I_{j,D}\right],\label{eA1_02}
			\end{align}
		\end{subequations}
		(2) Compute the extrinsic mean/variance of $\mathbf{x}$:
		\begin{subequations} \label{eq:eA1}
			\begin{align}
			&v_{A}^{\rm post}=\frac{1}{N}\sum\limits_{j=1}^{N}v_{j,A}^{\rm post},\label{eA1_04}\\
			&v_B^{{\rm pri}}=v_{A\to B}^{\rm ext}=\left(\frac{1}{v_{A}^{\rm post}}-\frac{1} {v_A^{\rm pri}(t)}\right)^{-1},\label{eA1_05}\\
			&\mathbf{x}_{A\to B}^{\rm ext}=v_{A\to B}^{\rm ext}\left(\frac{\mathbf{F}\mathbf{z}_A^{\rm post}}{v_{A}^{\rm post}}-\frac{\mathbf{F}\mathbf{z}_A^{{\rm pri}}}{v_A^{\rm pri}(t)}\right)\label{eA1_06},\\
			&\mathbf{x}_{d,B}^{{\rm pri}}=\mathbf{x}_{d,A\to B}^{\rm ext}\label{eA1_09},
			\end{align}
		\end{subequations}
		\textbf{Module B:}\\
		(3) Compute the posteriori mean/variance of $\mathbf{s}_d$:
		\begin{subequations}
			\begin{align}
			&s_{j,B}^{\rm post}=\mathrm{E}\left[s_{j}\mid \bar{h}_{j}, x_{j,B}^{{\rm pri}}\right],~j\in{\mathcal{X}}_d,\label{eA1_07}\\
			&v_{j,B}^{\rm post}=\mathrm{var}\left[s_{j}\mid \bar{h}_{j}, x_{j,B}^{{\rm pri}}\right],~j\in{\mathcal{X}}_d,\label{eA1_08}
			\end{align}
		\end{subequations}
		(4) Compute the extrinsic mean/variance of $\mathbf{z}$:
		\begin{subequations} \label{eq:e0530}
			\begin{align}
			&v_{B}^{\rm post}=\frac{1}{N}\sum_{j\in{\mathcal{X}}_d}|\bar{h}_j|^2v_{j,B}^{\rm post}\label{e0530_07b},\\
			&v_A^{{\rm pri}}(t+1)=v_{B\to A}^{\rm ext}=\left(\frac{1}{v_{B}^{\rm post}}-\frac{1}{v_B^{{\rm pri}}}\right)^{-1},\label{e0411_08a}\\
			&x_{j,B}^{\rm post} = {\bar{h}_j}s_{j,B}^{\rm post},~j\in{\mathcal{X}}_d,\label{e0530_07a}\\
			&\mathbf{z}_A^{{\rm pri}}(t+1)=\mathbf{z}_{B\to A}^{\rm ext}=v_{B\to A}^{\rm ext}\left(\frac{\mathbf{F}^H\mathbf{x}_B^{\rm post}}{v_{B}^{\rm post}}-\frac{\mathbf{F}^H\mathbf{x}_B^{{\rm pri}}}{v_B^{{\rm pri}}}\right). \label{e0411_08b}
			\end{align}
		\end{subequations}
		\vspace{-17pt}
	}
\end{algorithm}
 
\section{Numerical Results}
Before proceeding to the OTA experiments, we present the numerical simulation results of the proposed GTurbo-based algorithms.
The system parameters are specified as follows.
Notably, the following parameter setups will be followed in the OTA experiments in Section V-B.
\begin{itemize}
	\item  The number of orthogonal subcarriers $N$ is $2048$, and the number of subcarriers used for data transmission $N_d$ is $1186$.\footnote{For the transmission of 16-QAM symbols, only 1185 of them are actually used.}
	\item The turbo code is adopted for the channel coding technique, where the interleaver rule set based on \cite{r3}, and the recursive systematic convolutional encoders represented by [13, 15] with the constraint length of 4. 
	\item Uniform mid-rise quantization is adopted by each real-value quantizer $\mathcal{Q}(\cdot)$. The thresholds are set as $r_0=-\infty$, $r_b=(-2^{B-1}+b)\Delta$ for $b=1,2,\cdots,2^B-1$, and $r_{2^B}=\infty$, whereas the quantized outputs are set as $c_b=\left(-\frac{2^{B}+1}{2}+b\right)\Delta$ for $b=1,2,\cdots,2^B$, where quantization step size $\Delta=\sqrt{P_q}\Delta_B$ and $P_q=g_{\rm{AGC}}P_r/2$ denotes the input power of $\mathcal{Q}(\cdot)$.
	The values of $\Delta_B$ are typically selected to minimize the quantization MSEs, which are given by \cite{Max-1960TIT} and listed in Table I.
	\begin{table}[!ht]
		\renewcommand{\arraystretch}{1}
		\renewcommand{\arrayrulewidth}{1pt}
		\centering
		\caption{$\Delta_B$ for different  quantization bits $B$\label{T5}}
		\textcolor{black}{%
		\begin{tabular}{c c c c c c}
			\hline
			$B$ & 1 & 2 & 3 & 4 & 5\\
			\hline
			$\Delta_B$ & $\sqrt{{8}/{\pi}}$  &  0.9957  &  0.5860  &  0.3352  & 0.1881\\
			\hline
		\end{tabular}}
	\end{table}
	\item The entries of the pilot vector $\mathbf{p}_d$ are drawn independently from the equiprobable 4-QAM constellation. In addition, the entries of $\mathbf{p}_d$ and $\mathbf{s}_d$ are normalized, such that ${\rm E}[|p_j|^2] = 1$ and ${\rm E}[|s_j|^2] = 1$ for $j\in\mathcal{X}_d$.
\end{itemize}

In this section, we set the number of channel taps $L=4$, with power profiles $0$, $-7$, $-12$, and $-18$ dB for each tap.
We conduct Monte Carlo simulations of 1,000 independent channel realizations to obtain the performance metrics to be evaluated.
Imperfections in parameters $1/P_r$, $\sigma ^2$ and $L$ will be simulated by imposing certain mismatches on these parameters.
In particular, we artificially impose an error that is uniformly distributed in [-30\%, \, 30\%] on parameters $1/P_r$ and $\sigma ^2$ to obtain $g_{\rm{AGC}}$ and $\widehat {{\sigma ^2}}$, in addition, we let $\hat{L}$ = 6. 
These mismatches are set to simulate the practical situation in which exact parameters $1/P_r$, $\sigma ^2$ and $L$ are unavailable.
The influence of channel normalization in \eqref{Channel_norm} will also be discussed in addition to the imperfect parameters.
The performance under the perfect parameters, i.e., $g_{\rm{AGC}}= 1/P_r$, $\widehat {{\sigma ^2}}=\sigma ^2$, and $\hat{L}=4$, serves as the benchmark.
The following methods will be used later as benchmarks for performance comparison:
\begin{itemize}
	\item Conventional approach: the procedures described in section II that ignore quantization effect.
	\item Bussgang based method: the procedures described in section II based on Bussgang linearization for \eqref{DET_Prob_Form} and \eqref{CE_Prob_Form} in the same way of \cite[section IV-A]{Heath-2018Arxiv}.
	\item GAMP-based methods: EM-GM-GAMP \cite{Vila-2013TSP} for channel estimation and GAMP \cite[Algorithm 2]{Rangan-2010Arxiv} for detection.
	\item FASTA-based \footnote{In our simulation, we utilize part of the code downloaded from \url{https://github.com/quantizedmassivemimo/mu_mimo_ofdm/} and also FASTA toolbox \cite{Goldstein-2014Arxiv,Goldstein-2015Arxiv}.} maximum likelihood channel estimator and approximated MMSE detector \cite{Studer-2016TCOM}. 
\end{itemize}
In this section, we set the maximum iteration number as the stopping criteria for iterative methods.
We also mark the maximum iteration numbers we set in each figure, which provides clues about computational delay required by each iterative algorithm.

\begin{figure}[!t]
	\centering
	\subfloat[B = 1 bit]{\includegraphics[scale=0.6]{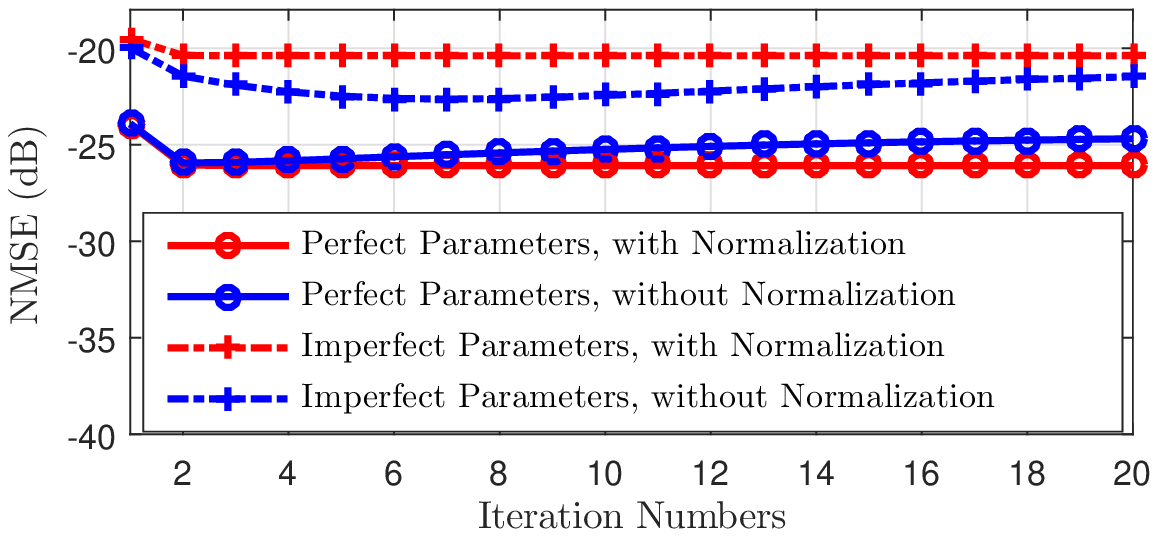}\label{F5A}}\\%
	\vspace{-10pt}
	\subfloat[B = 2 bits]{\includegraphics[scale=0.6]{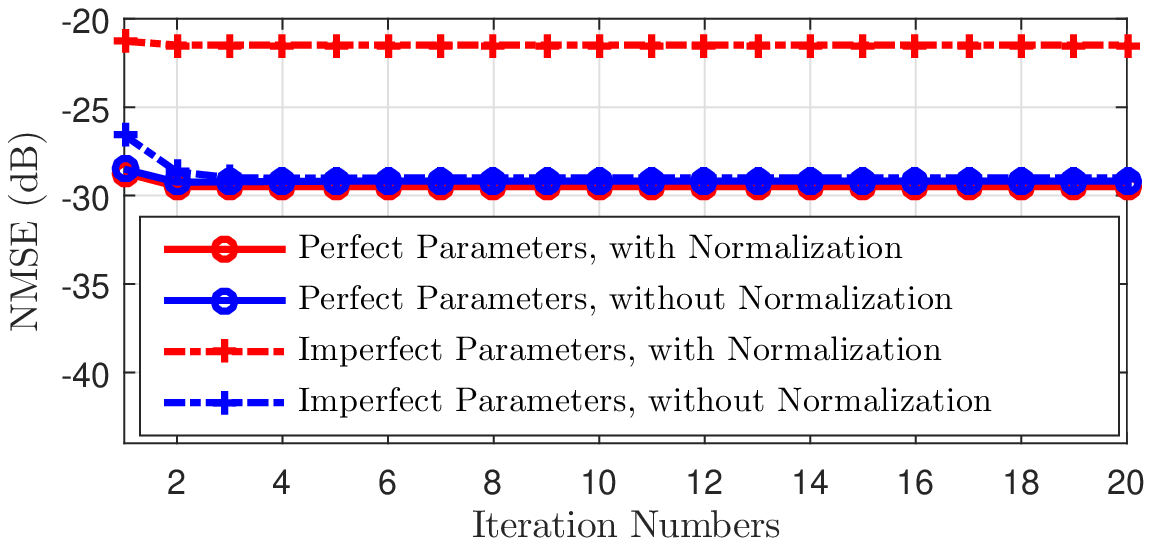}\label{F5B}}
	\caption{NMSEs versus iteration numbers of the proposed GTurbo-LMMSE channel estimator (i.e., Algorithm \ref{A1}) for an SNR of 12 dB under four conditions.\label{F5}}
\end{figure}

\begin{figure}[!t]
	\centering
	\subfloat[B = 1 bit]{\includegraphics[scale=0.4]{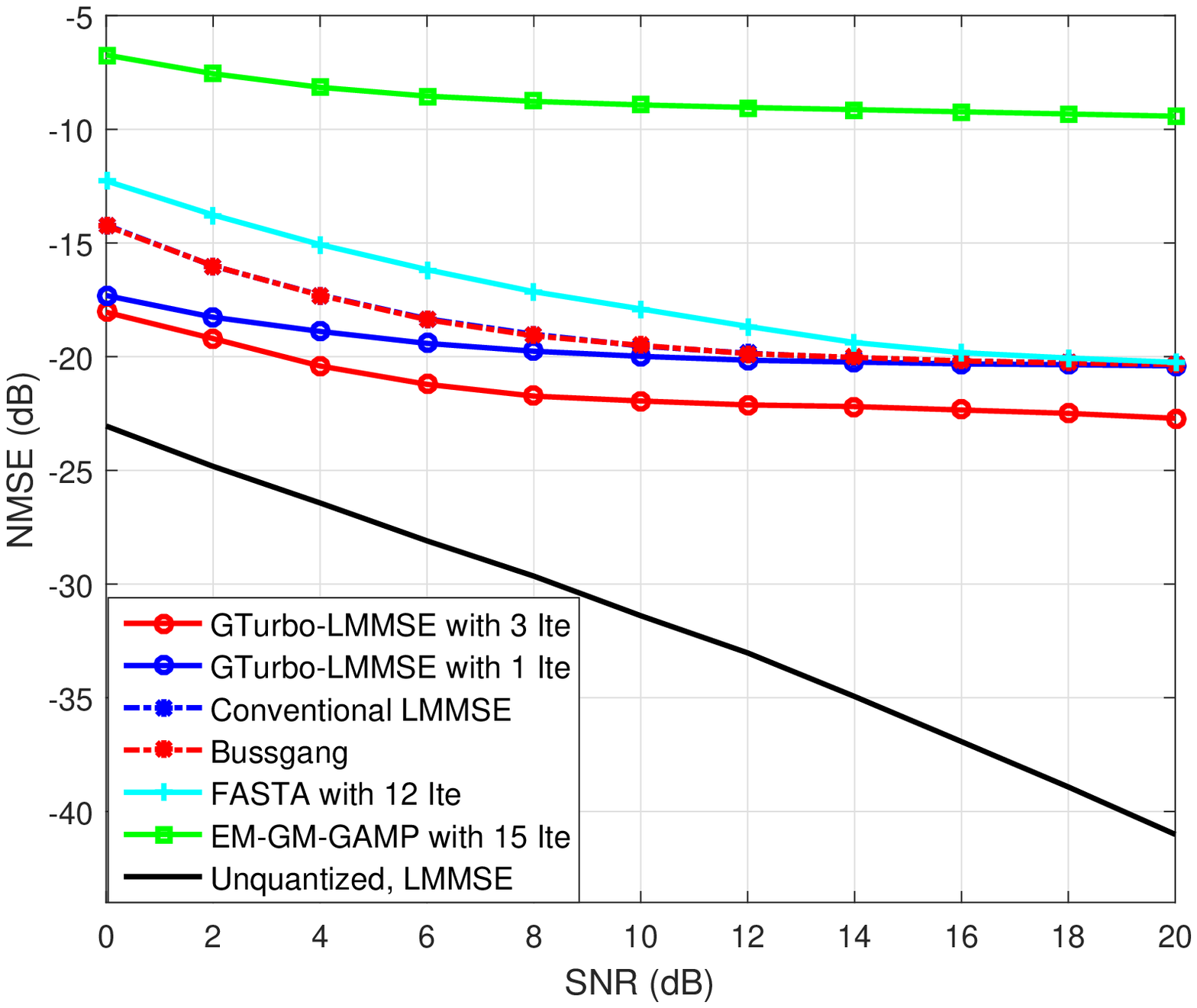}\label{F6A}}\\%
	\vspace{-10pt}
	\subfloat[B = 2 bits]{\includegraphics[scale=0.4]{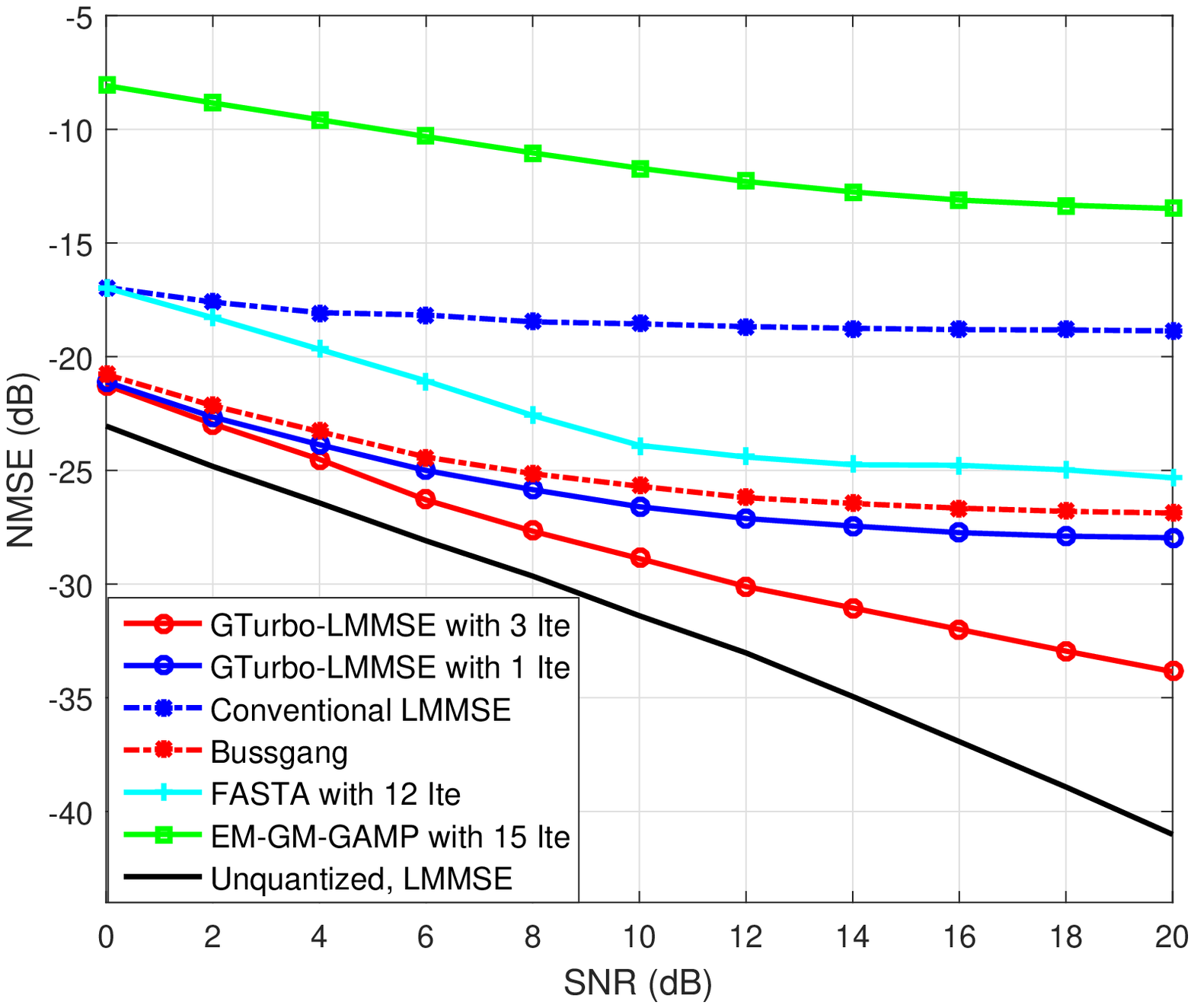}\label{F6B}}
	\caption{Comparison of the NMSE performances of various channel estimators when B = 1 bit and 2 bits under imperfect parameters.\label{F6}}
\end{figure}

We first present the normalized MSEs (NMSEs) of $\tilde{\mathbf{h}}_d$, which are defined as
\[\mathrm{NMSE}=\frac{\| {\tilde{\mathbf{h}}_d-\mathbf{h}}_d \|^2}{\left\| {\mathbf{h}}_d \right\|^2},\]
versus the iteration numbers of Algorithm 1. Performances are compared under four conditions produced by combining perfect or imperfect parameters, and whether channel normalization is performed.
Fig.~\ref{F5} shows the results under an SNR\footnote{We refer to SNR as the ratio of the average power of the noiseless received time-domain samples to the average noise power.} of 12dB and $B$ of 1 bit and 2 bits.
For $B = 1$ bit without channel normalization, the NMSEs are initially reduced and then increase again after a few iterations.
This is expected because the 1-bit ADCs implemented by signal zero-crossing comparators eliminate
the information regarding average channel gain $P_h$, which is difficult to recover using the proposed iterative procedure.
After phase information is recovered, the error in amplitude information leads to divergence.
Therefore, channel normalization is recommended when $B = 1$ bit to guarantee the numerical stability of the proposed algorithm.
Different observations can be made when $B = 2$ bits, under which NMSEs remains stable with only a minor fluctuation under all four conditions after a few iterations.
This result indicates that the phase information and amplitude information of $\mathbf{h}_d$ can be properly recovered using the proposed method.
When we compare the NMSE performances, better result is observed in the case of 2-bit quantization without channel normalization with imperfect parameters, which can also be observed when $B=$ 1 bit, whereas the opposite result is observed with perfect parameters.
This finding implies the inaccuracy of $\hat{P}_h$ estimated by \eqref{est_norm} from incorrect $g_{\rm{AGC}}$ and $\widehat {{\sigma ^2}}$, while the proposed algorithm exhibits strong signal recovery capability.
Hence, with $B=2$ bits or more, channel normalization is not preferable in the application of Algorithm 1 to practical systems.
Fig.~\ref{F5} also shows that a satisfactory channel estimate can be obtained with less than five iterations in all cases.

Fig.~\ref{F6} compares the NMSE performance of different algorithms under imperfect parameters.
The performance with full-resolution quantization is also included as a lower bound.
Evidently, the proposed algorithm outperforms its counterparts, which can be attributed to its capacity to update the linear approximation of the Q-OFDM system in \eqref{appro_AWGN} through iterations.
A channel estimate is produced (and can also be updated) based on this approximation.
We find from Fig.~\ref{F6} that the Bussgang-based method achieves a similar NMSE performance (with minimal performance degradation) to that of the proposed GTurbo-LMMSE channel estimator with one iteration.
EM-GM-GAMP does not work efficiently because it applies i.i.d. Gaussian mixture model to approximate the a priori distribution of each element of frequency-domain channel response, but the real channel responses are correlated.
The results in Fig.~\ref{F6} also indicate that when $B = 2$ bits, the gap between the NMSE performance of the proposed channel estimator and that of the LMMSE channel estimator under full-resolution quantization is minimal.

\begin{figure}[!t]
	\centering
	\subfloat[B = 1 bit, 4-QAM]{\includegraphics[scale=0.6]{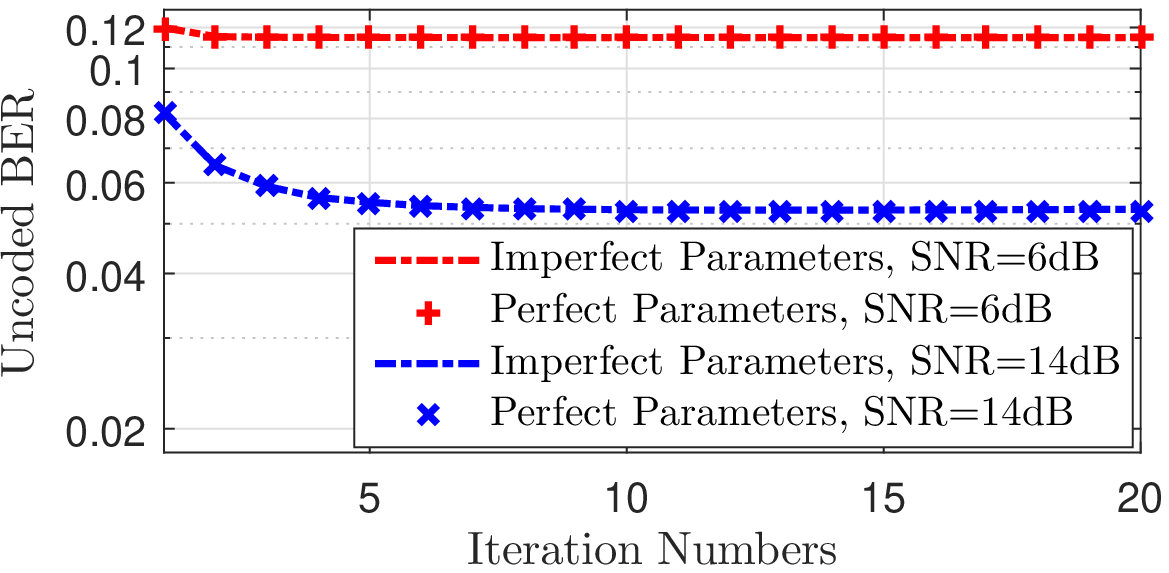}\label{F15A}}\\%
	\vspace{-10pt}
	\subfloat[B = 2 bits, 16-QAM]{\includegraphics[scale=0.6]{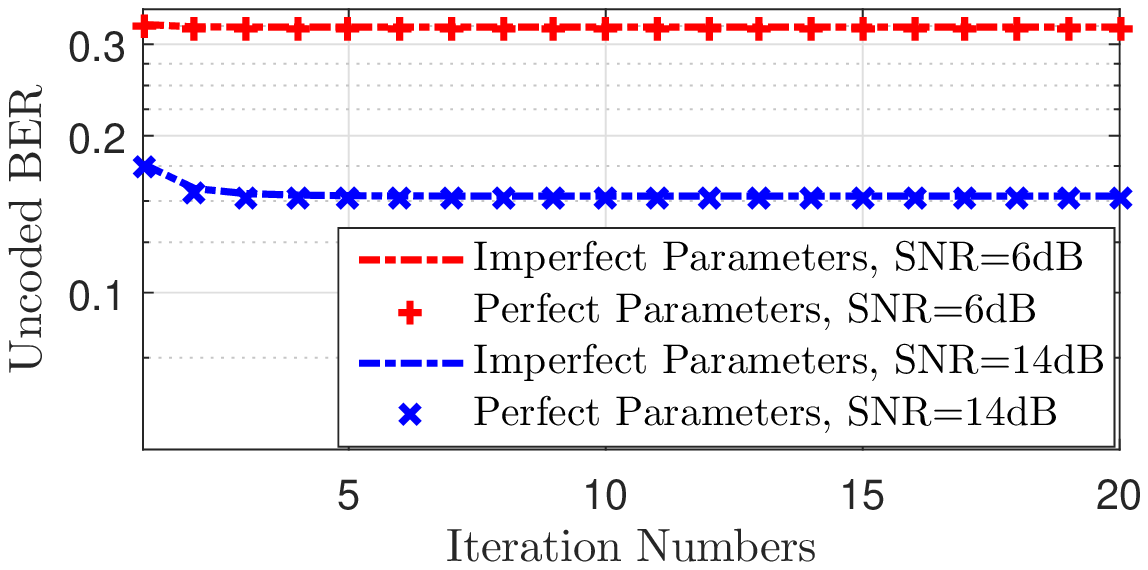}\label{F15B}}
	\caption{Uncoded BERs versus iteration numbers of the GTurbo-based detector (i.e., Algorithm \ref{A2}) under estimated channel responses.\label{F15}}
\end{figure}

\begin{figure}[!t]
	\centering
	\subfloat[B = 1 bit, 4-QAM]{\includegraphics[scale=0.48]{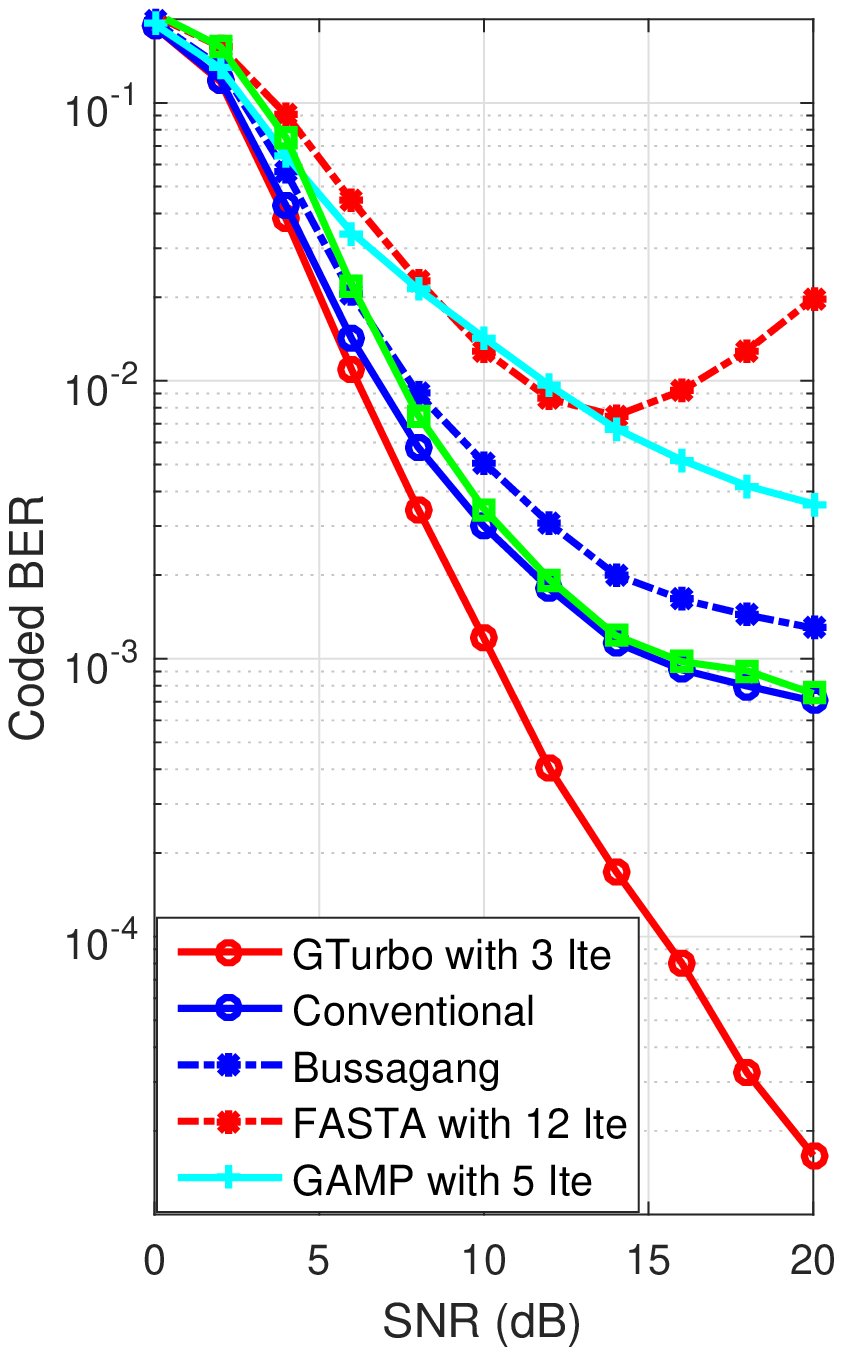}\label{F7A}}
	\subfloat[B = 2 bits, 16-QAM]{\includegraphics[scale=0.48]{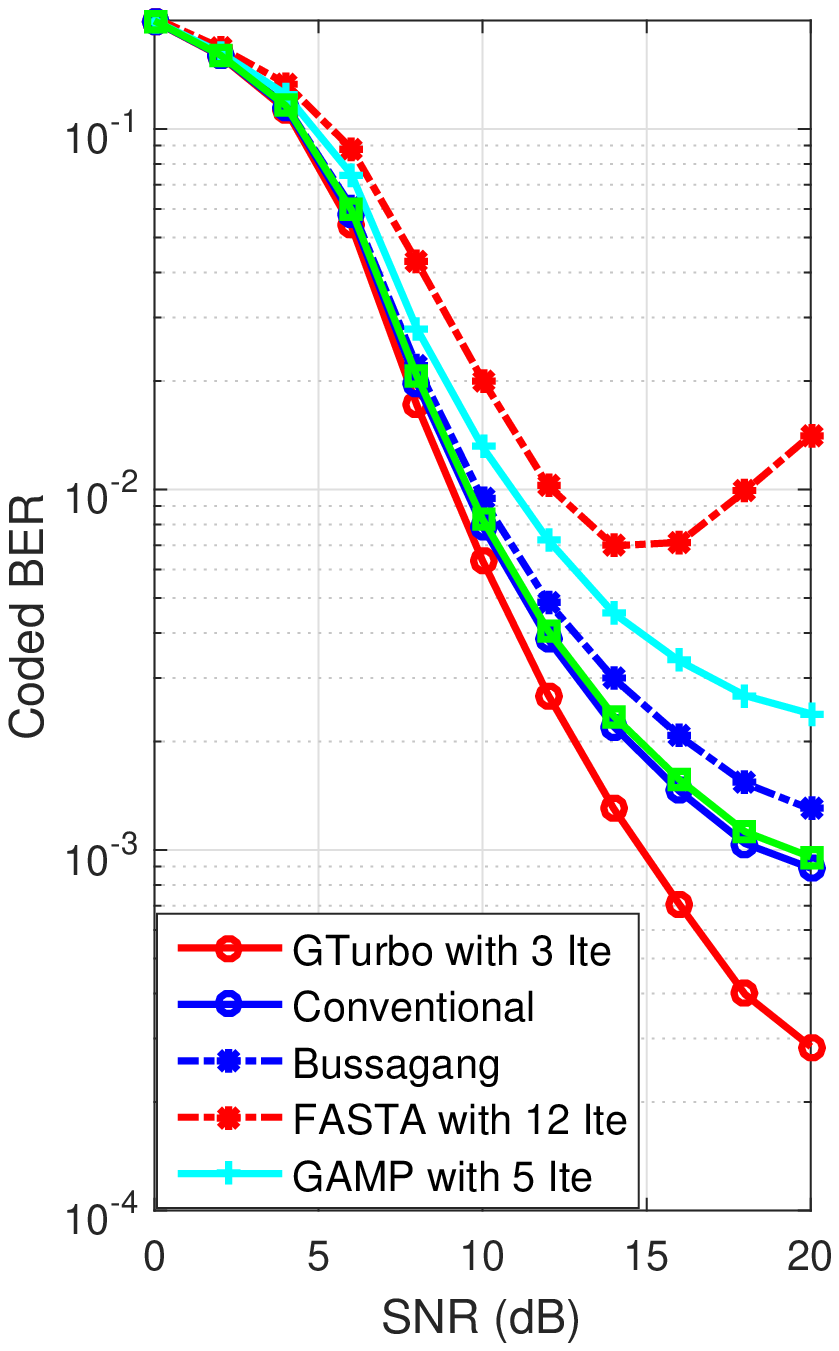}\label{F7B}}\\
	\caption{Comparison of data detection performances of various detectors when B = 1 bit and 2 bits under imperfect parameters and estimated channel responses.\label{F7}}
\end{figure}

We next evaluate and compare the information recovery performances of various algorithms under the estimated channel responses.
Fig. \ref{F15} shows the uncoded bit error rate (BER) of Algorithm 2 versus the iteration numbers under $\bar{\mathbf{h}}_d$ estimated by Algorithm 1.
At lower SNR (6 dB) and higher SNR (14 dB), and with perfect or imperfect parameters, the uncoded BERs are reduced firstly and then remain stable after a few iterations.
At most five iterations (typically 2 or 3 iterations are enough) are adequate for the algorithm to converge. 
In addition, performance loss from imperfect parameters is negligible.
This result justifies the robust signal recovery capability of the GTurbo-based algorithms.
In Figs. \ref{F7A} and \ref{F7B}, coded BER performances are compared.
We set the code rate to be 1/2 in this section to prevent the information recovery capabilities of the algorithms from being hidden by the strong channel code.
However, we set the code rate to be 1/3 to enable reliable communication in our prototyping system shown in section V-B.
From Figs. \ref{F7A} and \ref{F7B}, we have a similar finding that the Bussgang-based method has a similar coded BER performance (with minimal performance degradation) to that of the GTurbo-based detector with one iteration. 
In addition, we find that GTurbo is superior to GAMP because GAMP is derived for problems with linear transformation matrices that comprise i.i.d. Gaussian elements rather than orthogonal ones.
The proposed schemes also outperforms FASTA-based algorithms, given that the latter one assumes that the entries of $\mathbf{s}$ follow i.i.d. circularly symmetric complex Gaussian distribution instead of their true ones.
When computing LLR, the FASTA-based method cannot produce the variance information, i.e., $\beta_j $ in \eqref{Cal_LLR}, and approximate it by 1.
This approximation is inaccurate, especially under high SNR, which causes the coded BER curves to go up after a certain SNR for the cases of $B$ = 1 bit and 2 bits.

\section{Receiver Design and Implementation}
In this section, we first introduce the proposed Q-OFDM receiver architecture and then conduct OTA experiments to verify the reliability of our proposed architecture. 

\subsection{Receiver Architecture}

\begin{figure}[!t]
	\centering
	\includegraphics[scale=0.37]{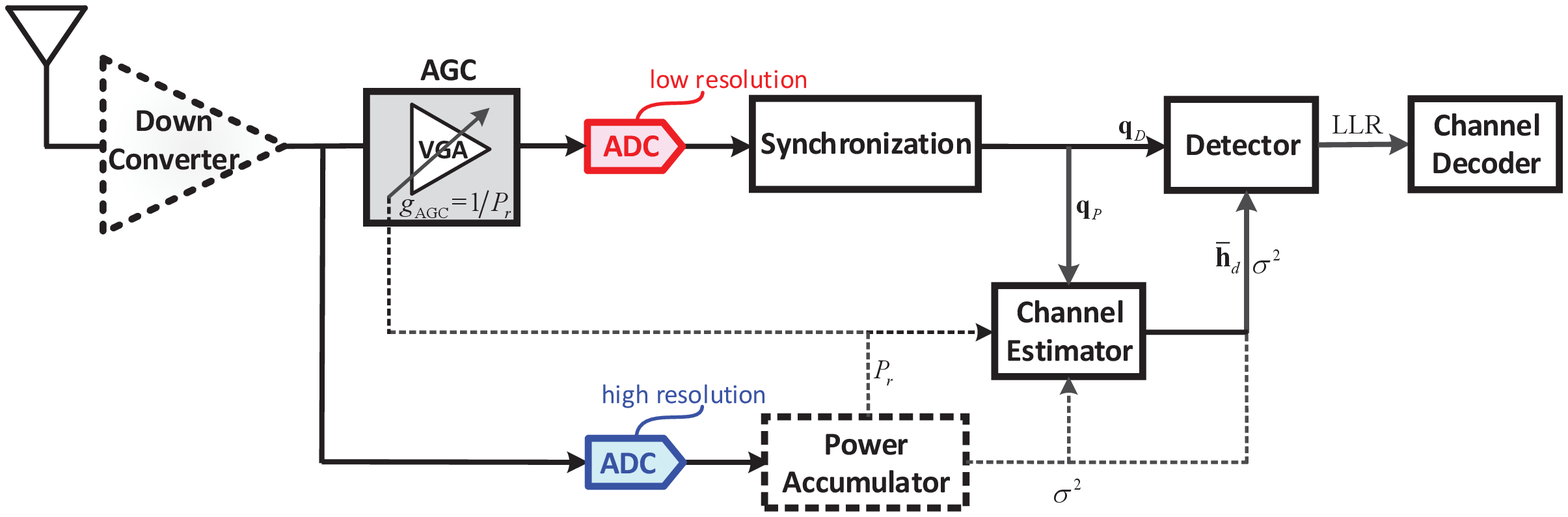}
	\caption{Q-OFDM receiver architecture.\label{F22}}
\end{figure}

Additional processing modules are required to guarantee that the inference algorithms proposed in Section III can be successfully implemented in practice, including AGC, signal/noise power estimator, and synchronization module.
Correspondingly, we propose the Q-OFDM receiver architecture shown in Fig. \ref{F22}.
Once the RF signal is received by the antenna from the wireless channel, the signal is down-converted into the analog baseband for subsequent processing.
The down converter module typically consists of low-noise amplifiers, local oscillators, filters, and mixers.
The received analog baseband signal is divided into two streams to be sampled and quantized in parallel with different sampling rates and quantization precisions for different purposes.
One stream is quantized with ultra-low resolution (e.g., 1$-$2 bits) and sampled with frequency $F_s$, from which synchronization searching, channel estimation, data detection and channel decoding are performed according to the adopted frame structure.
Another stream is quantized with high precision but sampled with a considerably lower sampling frequency (tens or hundreds times lower than $F_s$), which is used by the power accumulator
to assist in AGC, noise signal/noise estimation, and channel amplitude information recovery under 1-bit quantization.
We then elaborate how the power accumulator and synchronization module work.

\emph{1) Power Accumulator}:
The main purpose of this accumulator is to measure the average power of the high-resolution quantized samples within specific intervals.
The average power of the received signal $P_r$ is measured as a necessary parameter in the receiver.
In particular, we take the reciprocal of $P_r$ as the gain of the variable gain amplifier in the AGC unit. Hence, ${{{{\rm{g}}_{{\rm{AGC}}}}{\rm{ = 1}}} \mathord{\left/
		{\vphantom {{{{\rm{g}}_{{\rm{AGC}}}}{\rm{ = 1}}} {{P_r}}}} \right.
		\kern-\nulldelimiterspace} {{P_r}}}$.
This step enables us to adjust the amplitude of the baseband signal within a certain range to suit the fixed thresholds of the ADC.
For 1-bit quantization, the AGC unit becomes unnecessary.
However, $P_r$ is still a requisite parameter for the channel norm recovery shown in \eqref{est_norm},
which helps improve channel estimation performance as previously discussed and demonstrated.
In addition, noise power $\widehat {{\sigma ^2}}$ is another necessary parameter of the proposed algorithms.
Noise power estimation is conducted by assigning a few zero-padding OFDM symbols to each frame and measuring the average power of these OFDM symbols at the receiver.
The power accumulator measures noise power by using the high-resolution quantized samples of these OFDM symbols.

\begin{Remark}
	The power consumption of the assistant high-resolution ADC will be considerably reduced by decreasing its sampling frequency as much as possible, thereby
	leading to a reduction in the precision of signal/noise power estimation.
	The variations of channel gain and noise power are relatively slow.
	Therefore, we can compensate for the precision reduction by reasonably prolonging the time of the power accumulator.\hfill\ensuremath{\blacksquare}
\end{Remark}

\begin{figure}[!t]
	\centering
	\subfloat[B = 1 bit]{\includegraphics[scale=0.5]{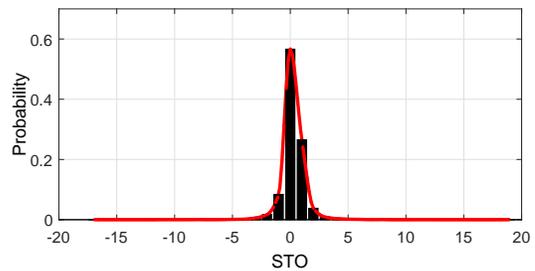}\label{F11A}}\\%
	\vspace{-10pt}
	\subfloat[B = 2 bits]{\includegraphics[scale=0.5]{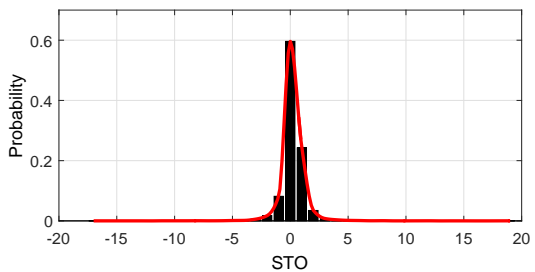}\label{F11B}}
	\caption{Distribution of the STO that corresponds to the exact synchronization position.\label{F11}}
\end{figure}

\emph{2) Synchronization}:
We perform the symbol-timing synchronization by following the correlation-based synchronization utilized in many extant communication systems to precisely identify the starting point of each frame.
We select the Zadoff Chu (ZC) sequence as the PSS because of its low cross-correlation and high auto-correlation.
For the OFDM symbols in each frame arranged to transmit PSS, the ZC sequence is placed at $N_s=62$ subcarriers corresponding to frequency $-\frac{N_s}{2}\frac{F_s}{N},\ldots,-\frac{F_s}{N},\frac{F_s}{N},\ldots,\frac{N_s}{2}\frac{F_s}{N}$, whereas other subcarriers are padded with zero to facilitate accurate synchronization.
We denote the vector of the frequency-domain OFDM symbol that contains the ZC sequence as ${\mathbf{t}\in\mathbb{C}^{N\times 1}}$ with ${t_j=0}$ for ${j\in\mathcal{X}\backslash\mathcal{X}_s}$, where $\mathcal{X}_s$ denotes the index subset of the $N_s$ subcarriers used for transmitting PSS.
We construct the reference sequence $\mathbf{t}_r$ by first transferring the frequency-domain OFDM symbol that transmits PSS into the time-domain which yields $\mathbf{\tilde{t}}=\mathbf{F}^H\mathbf{t}$, and then normalizing the power of the resulting sequence which yields $\mathbf{t}_r=\mathbf{\tilde{t}}/\|\mathbf{\tilde{t}}\|$.
Synchronization search is conducted by calculating the cross-correlation between the received quantized sequence (denoted by $\mathbf{q}_r$) and a reference sequence, namely, $\mathbf{q}^H_r\mathbf{t}_r$. 
The starting point of each frame is identified by determining the highest cross-correlation among a few adjacent symbol periods, which follows the same mechanism as that in full-resolution OFDM systems.
Once the starting point of each frame is identified, the CP of each OFDM symbol is removed, and the quantized time-domain OFDM symbols are inputted into their corresponding processing modules, namely, the channel estimator or data detector, by using the algorithms presented in Section III.

We conduct numerical simulation to examine whether the correlation-based synchronization is still valid for the coarsely quantized received signal.
Then, we display the probability distribution of the symbol time offsets (STOs), which are measured by the relative offset in the number of time-domain samples of the synchronization position that corresponds to the exact starting point of each frame  under an SNR of 15 dB and a quantization precision of 1$-$2 bits, as shown in Fig. \ref{F11}.
We draw the red curve to approximately depict the profile of the STO distribution.
The profile shows that the exact synchronization position can be identified with a large probability.
In particular, the probability of the exact synchronization is approximately 0.55 and 0.60 under the quantization precision of 1 bit and 2 bits, respectively.
Moreover, STO is retained within a range of $-4$ to $4$ with the highest probability, which is an acceptable synchronization error range.
We can conclude from the above preceding results that the correlation-based synchronization technique is still reliable for the received samples with low-resolution quantization.

\subsection{OTA Experiments}
\begin{figure}[!t]
	\centering
	\includegraphics[scale=0.4]{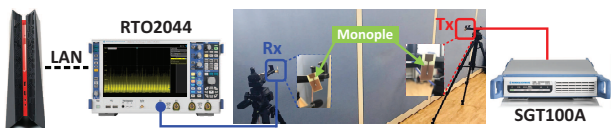}
	\caption{Picture of the proposed prototyping system.\label{F9}}
\end{figure}

\begin{figure}[!t]
	\centering
	\includegraphics[scale=0.45]{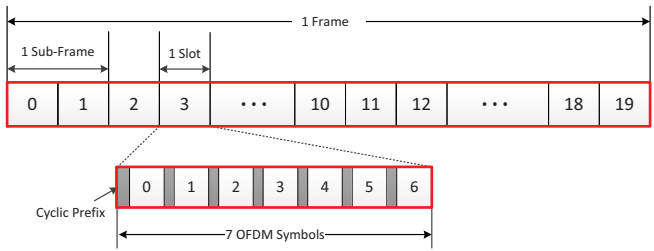}
	\caption{Frame structure\label{F1}}
\end{figure}

We then establish a proof-of-concept prototyping system based on the proposed architecture presented in the previous subsection (Fig. \ref{F9}).
The transmitter and the receiver are set up with a separate baseband and RF processing unit.
Baseband processing is performed through the software programmed in the host computer.
The corresponding procedures are specified in Sections III and V-A.
RF processing is conducted using the Rohde \& Schwarz equipments.
In particular, we use SGT100A for the transmitter and RTO2044 for the receiver.
The output baseband samples of RTO2044 are quantized by a high-resolution ADC.
Thus, coarse quantization is performed virtually through the software before further processing.
This setup enables us to obtain certain reference parameters for subsequent performance evaluations.
Data exchange between the baseband processing software and RF processing components is performed through off-the-shelf high-speed interfaces.

The frame structure used in our prototyping system is illustrated in Fig. \ref{F1}.
This structure is designed based on the frame structure of the LTE system, which can be flexibly reconfigured for future extensions.
An entire radio frame consists of 10 subframes, and each subframe has 2 consecutive slots.
In each slot, 7 OFDM symbols  are transmitted.
In our system, $L_{\rm cp}$ is set to 144 for OFDM symbols 1$-$6, and 160 for OFDM symbol 0 in each slot.
For each frame, slots 2$-$19 are dedicated for data transmission, in which the predefined pilot sequence is transmitted via the first OFDM symbol of these slots for channel estimation, whereas the remaining 6 OFDM symbols are used for transmitting data sequences.
To establish synchronization between the transmitter and the receiver, OFDM symbol 6 of slots 0 to 10 are reserved for transmitting PSS.
The first OFDM symbol of slot 1 is padded by null symbols for noise power estimation.
\begin{figure}[!t]
	\centering
	\includegraphics[scale=0.58]{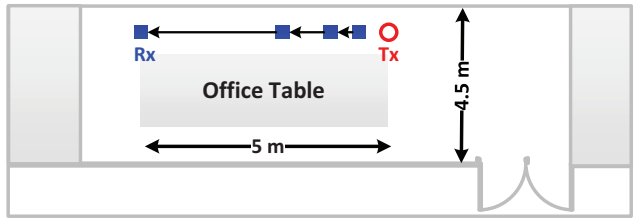}
	\caption{Experiment environment and equipment deployment.\label{F8}}
\end{figure}

The experiments are conducted in an indoor environment displayed in Fig. \ref{F8}.
The transmitter and the receiver are placed in a corridor beside an office table.  
We adjust their distance by fixing the transmitter and placing the receiver in different positions.
In our experiments, the carrier frequency is set to 3.5 GHz, and we set the bandwidth to be 100 MHz to support high data transmission with a constant frame duration.
The resulting system parameters are listed in Table \ref{T4}. 

\begin{table}[!t]
	\renewcommand{\arraystretch}{1.1}
	\centering
	\caption{System Parameters\label{T4}}
	\begin{tabular}{l c}\hline
		\textbf{Parameter} & \textbf{Value}\\
		\hline
		System Bandwidth & 100 MHz \\
		Sampling Rate $F_s$ & 153.6 MHz\\
		Subcarrier Interval & 75 KHz\\
		Frame Duration & 2 ms\\
		Subframe Duration & 0.2 ms\\
		OFDM symbol Duration (without CP) & 13.33 $\mu$s\\
		\hline
	\end{tabular}
\end{table}

\begin{table*}[!t]
	\renewcommand{\arraystretch}{1.3}
	\centering
	\caption{B = 1 bit, 4-QAM\label{T2}}
	\begin{tabular}{c c c c c c c}\hline\hline
		\textbf{Transmitted Power} & \textbf{Distance} & \textbf{SNR} & \textbf{MSE} & \multirow{2}*{\textbf{Uncoded BER}} & \multirow{2}*{\textbf{Coded BER}} & \multirow{2}*{\textbf{PER}}\\
		\textbf{(dBm)} & \textbf{(cm)} & \textbf{(dB)} & \textbf{(dB)}\\
		\hline
		\multirow{4}*{-18}
		& 50 & 12.6768 & -21.1096 & 0.0385 & 0 & 0\\
		& 100 & 7.8791 & -17.1139 & 0.0762 & 0 & 0\\
		& 200 & 5.4760 & -15.2942 & 0.1610 & 7.6436$\times10^{-5}$ & 0.0230\\
		& 250 & 3.9743 & -14.0242 & 0.1820 & 0.0017 & 0.2030\\
		\hline
		\multirow{2}*{-15}
		& 200 & 5.8339 & -15.5368 & 0.1574 & 5.6588$\times10^{-5}$ & 0.0080\\
		& 250 & 5.5803 & -15.4472 & 0.1644 & 1.0093e$\times10^{-4}$ & 0.0310\\
		\hline\hline
	\end{tabular}
\end{table*}

\begin{table*}[!t]
	\renewcommand{\arraystretch}{1.3}
	\centering
	\caption{B = 2 bits, 16-QAM\label{T3}}
	\begin{tabular}{c c c c c c c}\hline\hline
		\textbf{Transmitted Power} & \textbf{Distance} & \textbf{SNR} & \textbf{MSE} & \multirow{2}*{\textbf{Uncoded BER}} & \multirow{2}*{\textbf{Coded BER}} & \multirow{2}*{\textbf{PER}}\\
		\textbf{(dBm)} & \textbf{(cm)} & \textbf{(dB)} & \textbf{(dB)}\\
		\hline
		\multirow{4}*{-18}
		& 50 & 12.6867 & -22.8611 & 0.1128 & 0 & 0\\
		& 100 & 7.9481 & -18.5790 & 0.1633 & 0 & 0\\
		& 200 & 6.3335 & -17.2023 & 0.2022 & 0.0108 & 0.5160\\
		& 250 & 4.0372 & -15.1170 & 0.2455 & 0.1791 & 1\\
		\hline
		\multirow{2}*{-15}
		& 200 & 6.7478 & -17.6764 & 0.1879 & 1.4032$\times10^{-4}$ & 0.0269\\
		& 250 & 6.5489 & -17.3654 & 0.1968 & 0.0020 & 0.2405\\
		\hline\hline
	\end{tabular}\\
\end{table*}

Experiments are conducted to evaluate the error rate performance of the proposed prototyping system under different quantization precisions and modulation schemes, including 1 bit for 4-QAM and 2 bits for 16-QAM.
Various values of average SNRs are generated by changing the transmission power and distance between the transmitter and the receiver.
The performance metrics to be evaluated include the MSE of the estimated channel parameter, coded and uncoded BER, and PER of the information bits in each code word.
The average SNRs and these metrics are measured from 10,000 received frames and obtained by calculating the average of these measured results.
We use the LMMSE channel estimate from the high-resolution received signal as the reference when calculating the MSE of the estimated channel parameters.
The experiment results are presented in Tables \ref{T2} and \ref{T3}, which show the measured performance metrics when transmitting 4-QAM with 1-bit quantization and 16-QAM with 2-bit quantization, respectively.
We can conclude that reliable data transmission can be guaranteed when SNR is higher than approximately 6 dB for both cases.

\section{Discussions}
\subsection{Implementation Issues}
In this subsection, we discuss the implementation issues of GTurbo-based algorithms. 
Compared with linear receivers, which involve only one matrix-vector multiplication, GTurbo-based algorithms involve many challenges in their hardware-friendly implementation. 
These challenges include various nonlinear computations in Module A, several matrix-vector multiplications, and the need for iteration. 
The rest of this subsection shows that these GTurbo-based algorithms can be hardware-friendly and efficiently implemented. 

First, we elaborate on the implementation of those nonlinear computations and matrix-vector multiplications. 
The computations of the posteriori expectation and variance in \eqref{eIII_01} and \eqref{eIII_02} require deriving the integral of the Gaussian distribution function $\phi(x)$, which poses the largest bottleneck in algorithm implementation.
Hardware-friendly approximations for $\phi(x)$ and $Q(x)$ were proposed in \cite{CZhang-2016SiPS}, which also facilitate the nonlinear operations in \eqref{eIII_05} and \eqref{eIII_06}.
The matrix multiplications in \eqref{ext_mean_A1}, \eqref{ext_var_B}, \eqref{eA1_06} and \eqref{e0411_08b} can be implemented using fast Fourier transform processors with computational complexity $\mathcal{O}(N\log_2N)$.
Our approximation for the weight matrix of the LMMSE channel estimator in \eqref{LMMSE_Mat} enables us to compute this matrix off-line and store the result in the memory, thus we can avoid the online computation of high-dimensional matrix inversion.
We can also observe that the two algorithms demonstrate the same operations in Module A, which indicates that they can share the identical circuit module when implementing the proposed architecture in practice.
Interestingly, from Algorithms 1 and 2, we find that the proposed GTurbo-based methods contain the operation of the conventional method, specifically, \eqref{LMMSE_denoiser} with $\mathbf{x}_{B}^{{\rm pri}}=\mathbf{F}\mathbf{q}_P$ in Algorithm 1, and \eqref{Cal_LLR} with $\mathbf{x}_{B}^{{\rm pri}}=\mathbf{F}\mathbf{q}_D$ and $v_{B}^{{\rm pri}}$ being the estimate of noise power (and quantization error) in algorithm 2. 
This feature provides the flexibility to reduce the iterative procedures to the conventional operations by disabling other operations and allowing only one iteration in the condition that iteration cannot yield sufficient performance improvement.

The hardware-friendly implementation of the entire iterative procedure can be achieved by applying the pipelined and folding hardware architecture proposed in \cite{CZhang-2016SiPS}. 
Moreover, Section IV presents that the iterative process can converge after only two or three iterations, thereby preventing high computational delay.
These facts make efficient and real-time implementation of algorithms feasible. 
The simulation results in \cite{CZhang-2016SiPS} demonstrate that the fixed-point setting combined with Q-function approximation causes only a minimal performance degeneration compared with the original floating-point simulation.
Furthermore, our MIMO-OFDM prototyping \cite{Gao-2018ACCESS} verifies the capacity of multi-core processors in implementing high-speed signal processing and complicated operations, which implies the potential of using multi-core processors to effectively implement advanced signal-processing techniques (like GTurbo) in practice.

\subsection{Future Works}

In this subsection, we discuss our initial ideas about the extensions to multi-user and multi-antenna cases. 
The following are the two potential directions for these extensions. 
The first direction involves aligning the proposed algorithms with advanced beamforming techniques. 
The second direction involves developing novel techniques that are specialized for scenarios associated with multiple received streams.
The first direction is promising for the systems in which we can generate high-directional beams between transmitters and receivers. 
Here, we use the uplink transmission as an example. 
Narrow beams are steered via a specific beamforming technique for the formation of high-directional spatial links between different users and the base station (BS). 
Through such beamforming techniques aligned with proper user selection, the multi-antenna and multi-user systems can be regarded as a bank of parallel single-stream transmissions. 
Therefore, the proposed techniques can be directly used at the BS side for each individual spatial link. 
Moreover, the key point, which requires our future works on this direction, is the design of reliable beamforming techniques.	
	 
However, for scenarios with a wide angle spread or when beams steered for different links overlap, the previous idea is ineffective. 
Novel algorithm frameworks dedicated to mitigating nonlinear distortion and multi-user interference must therefore be developed. 
The inference problems in Q-OFDM systems with multiple users and multiple reveived streams involve similar formulation as \eqref{DET_Prob_Form} with a generic linear transformation matrix. 
The extended algorithm framework is proposed in \cite{He-2018ISIT} by constraining the estimation into the linear space characterized by a linear transformation matrix, which is reduced to GTurbo when the linear transformation matrix is an orthogonal matrix.
This work exhibits potential for the detection of the single-user MIMO case.   
Subsequent researches on this new framework is currently ongoing on the following two aspects. 
First, the use of algorithm in \cite{He-2018ISIT} requires unitary precoding among all transmitted streams at the transmitter, which restricts its use to multi-user cases. 
Second, the efficient channel estimation method based on this new framework is being developed.

\section{Conclusion}
This study focuses on the development of a reliable Q-OFDM receiver architecture and the prototyping system implementation.
A novel Q-OFDM channel estimator was developed by integrating a type of robust linear OFDM channel estimator into the original GTurbo framework with extrinsic information derived to guarantee its convergence.
In addition, synchronization, AGC, and signal/noise power measurement were also properly considered.
We proposed an efficient Q-OFDM receiver architecture by combining the above issues with the proposed GTurbo-based algorithms, based on which we constructed a Q-OFDM prototyping system. 
OTA experiments are conducted to justify the reliability of the proposed architecture.
The following conclusions can be drawn from the results.
\begin{itemize}
	\item The proposed GTurbo-LMMSE channel estimator outperforms its counterparts in terms of NMSE performance.
	Performance under 2-bit quantization is very close to that of full-resolution quantization.
	In addition, additional recovery of amplitude information is required in the case of 1-bit quantization.
	
	\item The GTurbo-based detector, with support of the proposed GTurbo-LMMSE channel estimator, obtains performance gain in terms of coded BER compared with its counterparts. 

	\item Correlation-based synchronization still works well in the coarsely quantized cases.

	\item A communication system that supports reliable OFDM transmission with ultra-low resolution ADCs is achievable.	
\end{itemize}

\section*{Acknowledgment}
We would like to thank Prof. Yonina Eldar and Dr. Nir Shlezinger of Technion-Israel Institute of Technology for their great efforts and valuable suggestions on this work.

\ifCLASSOPTIONcaptionsoff
  \newpage
\fi



%

\bibliography{IEEEabrv,myBib}

%





\end{document}